\documentclass{JHEP3}

\usepackage{graphicx}
\usepackage{bm}
\usepackage{amsmath}
\usepackage{amstext}
\usepackage{amssymb}

\title{The Sommerfeld enhancement for dark matter with an excited state}

\author{Tracy R. Slatyer\\
Physics Department, Harvard University, Cambridge, MA 02138, USA\\
E-mail: \email{tslatyer@fas.harvard.edu}}

\abstract{
We present an analysis of the Sommerfeld enhancement to dark matter annihilation in the presence of an excited state, where the interaction inducing the enhancement is purely off-diagonal, such as in models of exciting or inelastic dark matter. We derive a simple and accurate semi-analytic approximation for the $s$-wave enhancement, which is valid provided the mass splitting between the ground and excited states is not too large, and discuss the cutoff of the enhancement for large mass splittings. We reproduce previously derived results in the appropriate limits, and demonstrate excellent agreement with numerical calculations of the enhancement. We show that the presence of an excited state leads to generically larger values of the Sommerfeld enhancement, larger resonances, and shifting of the resonances to lower mediator masses. Furthermore, in the presence of a mass splitting the enhancement is no longer a monotonic function of velocity: the enhancement where the kinetic energy is close to that required to excite the higher state can be up to twice as large as the enhancement at zero velocity.}


\preprint

\begin{document}

\section{Introduction}

In particle physics, perturbation theory is commonly employed to calculate annihilation and scattering cross sections, with higher-order terms in the perturbative expansion being neglected. Provided the theory is not strongly coupled, this is generally a good approximation for relativistic particles, but at low velocities and in the presence of a long-range force (classically, when the potential energy due to the long-range force is comparable to the particles' kinetic energy), the perturbative approach breaks down. In the nonrelativistic limit, the question of how the long-range potential modifies the cross section for short-range interactions can be formulated as a scattering problem in quantum mechanics, with significant modifications to the cross sections occuring when the particle wavefunctions are no longer well approximated by plane waves (so the Born expansion is not well-behaved). The deformation of the wavefunctions due to a Coulomb potential was calculated by Sommerfeld in \cite{sommerfeld}, yielding a $\sim 1/v$ enhancement to the cross section for short-range interactions (where the long-range behavior due to the potential can be factorized from the relevant short-range behavior).

Recent measurements of the positron and electron cosmic ray spectra by the PAMELA, ATIC, Fermi and H.E.S.S experiments have observed a rise in the positron fraction starting at $E \sim 10$ GeV and extending (at least) up to $E \sim 100$ GeV \cite{Adriani:2008zr}, as well as a broad excess in the total $e^+ + e^-$ spectrum extending from several hundred GeV to $\mathcal{O}(\mathrm{TeV})$ energies \cite{aticlatest, Abdo:2009zk, Aharonian:2009ah}. Observations by WMAP \cite{Bennett:2003bz,Hinshaw:2008kr} also suggest an excess in microwave emission from the inner Galaxy, termed the ``WMAP Haze'' \cite{Finkbeiner:2003im}, which has been attributed to synchrotron radiation from a new population of high energy electrons (with energies of tens to hundreds of GeV) \cite{Finkbeiner:2004us, Dobler:2007wv, Hooper:2007kb}. 

There has been considerable interest in these observations as possible signatures of dark matter (DM) annihilation or decay. If DM annihilation is responsible for the excesses, the annihilation cross section must exceed that required for a WIMP initially in thermal equilibrium to freeze out with the correct relic density, by 1-4 orders of magnitude depending on the dark matter mass. Furthermore, to avoid antiproton bounds from the PAMELA experiment \cite{Adriani:2008zq} and to generate sufficiently hard $e^+ e^-$ spectra, the dark matter should annihilate largely into leptons, pions, kaons and other light states (e.g. \cite{Cirelli:2008pk, Meade:2009iu}).

If dark matter is self-interacting, either via exchange of Standard Model gauge bosons or due to some novel interaction, then the nonperturbative ``Sommerfeld enhancement'' \emph{must} be taken into account when computing annihilation cross sections in the present-day Galactic halo. The importance of the Sommerfeld enhancement in the context of dark matter annihilation has been studied  by several authors \cite{Hisano:2003ec, Hisano:2004ds, Cirelli:2007xd, MarchRussell:2008yu}. Due to the velocity dependence of the Sommerfeld effect, and the presence of large resonances due to near-threshold bound states at low velocity (see e.g. \cite{MarchRussell:2008tu}), the annihilation cross section could potentially be greatly increased in the present-day Galactic halo ($\beta \sim 10^{-4}-10^{-3}$), without greatly perturbing the freezeout of annihilations at $\beta \sim 0.3$ (however, the Sommerfeld enhancement can modify the relic density at some level, particularly in the presence of resonances, see e.g. \cite{Dent:2009bv, Zavala:2009mi}). 

Furthermore, it was pointed out in \cite{Finkbeiner:2007kk,Cholis:2008vb} that the presence of light ($\lesssim$ GeV) force carriers coupling to the dark matter would naturally lead to unusual dark matter annihilation signatures, with zero or small branching ratios into heavy hadrons and gauge bosons, and hard spectra of leptons, pions and other light states. The dominant annihilation channel in such models would be annihilation of the dark matter into the new force carriers, which would then decay into kinematically accessible SM states. In particular, provided the mass of the new force carrier was less than twice the proton mass, no excess antiprotons would be produced. A new GeV-scale force in the dark sector could therefore naturally generate both a large annihilation cross section via Sommerfeld enhancement, and the observed $e^+ e^-$ spectra, without violating antiproton constraints \cite{ArkaniHamed:2008qn, Pospelov:2008jd}.

As noted in \cite{ArkaniHamed:2008qn}, if dark matter is charged under some new dark gauge group, then when the new gauge boson acquires an $\mathcal{O}$(GeV) mass the states in the dark matter multiplet can naturally be split from each other by a small ($\mathcal{O}$(MeV)) amount (this happens automatically, due to loops of the new dark gauge bosons, if the gauge group is non-Abelian \cite{ArkaniHamed:2008qn, Baumgart:2009tn}, but such mass splittings can also be naturally generated in Abelian models, e.g. \cite{Cheung:2009qd}). Furthermore, if the dark matter is a Majorana fermion or real scalar, its couplings with any vector boson must be off-diagonal, since it cannot carry conserved charge. Models of this type, where the dark matter can only scatter inelastically at tree level (that is, by exciting some slightly more massive state) are also motivated by the iDM \cite{Smith:2001hy, Chang:2008gd} and XDM \cite{Finkbeiner:2007kk} scenarios, which explain, respectively, the anomalies observed by the DAMA/LIBRA \cite{Bernabei:2008yi} and INTEGRAL/SPI \cite{Weidenspointner:2006nu} experiments.

Despite the fairly generic presence of these excited states, most recent analyses of the Sommerfeld enhancement in the context of the cosmic-ray anomalies have treated only the case of degenerate dark matter states, under the presumption that the introduction of an excited state does not significantly modify the enhancement to annihilation, unless the mass splitting is large enough to cause the enhancement to cut off (see \cite{ArkaniHamed:2008qn} for an argument to this effect) \footnote{The related problem of collisional excitation of these states, via scattering mediated by a long-range force, has been considered by several authors, e.g. \cite{Finkbeiner:2007kk,Chen:2009dm}.}. While it is true that the general qualitative features of the enhancement are mostly unchanged unless the mass splitting is quite large, modifications to the resonance structure occur for smaller mass splittings, and should be taken into account in any detailed analysis of the enhancement. However, the addition of even a single mass splitting adds an extra parameter to the problem, thus making numerical studies significantly more time-consuming; the numerical calculation of the enhancement can also become badly unstable in some parts of parameter space. 

In \S \ref{sec:deriv} we derive simple expressions to aid in computing the Sommerfeld enhancement in the general case where the $l$th partial wave dominates annihilation, for an arbitrary potential coupling $N$ states. In \S \ref{sec:parametrics} we discuss the general properties of the case where the dark matter has a single excited state and the interaction is purely off-diagonal, including the limit of zero mass splitting and a characterization of the parts of parameter space where the enhancement is negligible. In \S \ref{sec:approxsol} we derive an approximate semi-analytic solution for the Schr\"{o}dinger equation in the case of a two-state system with purely off-diagonal interaction, focusing on the case of $s$-wave annihilation. In \S \ref{sec:analysis} we present a simple analytic approximation for the Sommerfeld enhancement, strictly valid when the annihilation channels and matrix elements are the same for two particles in the excited state as in the ground state, and also applicable to the more general case (up to a simple rescaling), provided the enhancement is large. We employ this approximation to discuss the new features present in the case with a non-zero mass splitting, and confirm our results by numerical simulations.

\section{The Sommerfeld enhancement for a general $N$-state system}
\label{sec:deriv}

The Sommerfeld enhancement results from the distortion of the two-body wavefunction away from a plane wave, in the presence of a long-range potential. Provided that the range of the interaction responsible for annihilation is much shorter than the range of the potential, we can approximate the annihilation interaction as a delta function at the origin (in relative coordinates). Then the Sommerfeld-enhanced annihilation rate can be determined by first solving the scattering problem (for the chosen initial plane wave), and then contracting the state vector at the origin with the matrix describing the annihilation rate for the various states, which we will denote $\Gamma$. The fraction of scatterings involving transitions to the various states can be determined similarly, by solving the scattering problem for an initial plane wave purely in one two-body state, and then reading off the amplitudes of the (purely outgoing) partial waves in the other states.

In terms of Feynman diagrams, the Sommerfeld enhancement can be expressed as the result of resumming an infinite series of ladder diagrams. Several authors have shown that in the non-relativistic limit, this procedure is equivalent to solving the Schr\"{o}dinger equation with an appropriate (in general, matrix-valued) potential \cite{Hisano:2004ds, Iengo:2009ni, Cassel:2009wt}.

Let us assume a spherically symmetric, real-valued $N \times N$ matrix potential, and transform to relative coordinates. Then as usual, separation of variables yields the radial Schr\"{o}dinger equation for the two-particle wavefunction,
\begin{equation} \frac{d}{d r} \left(r^2 \frac{d \vec{R}_l(r)}{d r} \right) - \frac{m_\chi r^2}{\hbar^2} \left( \mathcal{V}(r) - m_\chi v^2 \right) \vec{R}_l(r) = l(l+1) \vec{R}_l(r),  \end{equation} 
for the $l$th partial wave, where $\vec{R_l}(r)$ is a regular real $N$-component vector function, $m_\chi$ is the dark matter mass (twice the reduced mass), and we incorporate any mass splitting terms into the potential matrix $\mathcal{V}$ (we assume throughout that any mass splittings are much smaller than the dark matter mass). There are $2 N$ linearly independent (LI) solutions to this equation in total, but the requirement that all components be regular at the origin fixes $N$ of the boundary conditions. Let us promote $R_l(r)$ to an $N \times N$ matrix, such that its columns are $N$ LI regular vector solutions $\vec{R}_l(r)$. The functions $R_l(r)_{i j}$ have the asymptotics as $r \rightarrow \infty$,
\begin{equation} (R_l(r))_{i j} \rightarrow \frac{1}{r} \sin \left(k_i r - \frac{1}{2} l \pi + (\delta_l)_{i j} \right) (n_l)_{i j}, \label{eq:largerbasis} \end{equation} 
where $(n_l)_{i j}$ and $(\delta_l)_{i j}$ are real numbers, and $k_i = \sqrt{(m_\chi v)^2 - m_\chi \mathcal{V}_{ii}(\infty)}/\hbar$, where we assume that $\mathcal{V}(\infty)$ is diagonal (since the elements of $\mathcal{V}$ that do not vanish as $r \rightarrow \infty$ correspond to mass terms, and we can choose to work in the basis of mass eigenstates). We will write the general solution to the Schr\"{o}dinger equation in the form $ \vec{\Psi}(r, \theta) = \sum_{l=0} P_l(\cos{\theta}) R_l(r) \vec{A}_l, $
where $\vec{A}_l$ is a complex $N$-vector giving the coefficients of the real vector functions in $R_l(r)$. 

The scattering solution to the multi-state Schr\"{o}dinger equation can be written,
\begin{equation} \Psi_n(r,\theta) = c_n e^{i k_n z} + f_n(r,\theta) e^{i k_n r} / r, \label{eq:scatteringsolution} \end{equation} 
where the $N$-vector $\vec{c}$ describes the initial incoming plane wave and $\vec{f}$ describes the outgoing scattered wave. Using the asymptotic expansion of $e^{i k z}$,
\begin{equation} e^{i k z} \rightarrow \frac{1}{2 i k r} \sum_l (2 l + 1) P_l(\cos \theta) ( e^{i k r} - (-1)^l e^{- i k r} ), \end{equation}
for each $l$ we can write
\begin{eqnarray} (\psi_n)_l & \rightarrow & c_n \frac{1}{2 i k_n r} (2 l + 1) P_l(\cos \theta) ( e^{i k_n r} - (-1)^l e^{- i k_n r} ) + f_n e^{i k_n r} / r \nonumber \\ & = & P_l (\cos \theta) \frac{1}{2 i r} \sum_{j} (e^{i \left(k_n r - \frac{1}{2} l \pi + (\delta_l)_{n j} \right)} - e^{- i \left(k_n r - \frac{1}{2} l \pi + (\delta_l)_{n j} \right)})  (n_l)_{n j} (A_l)_j. \end{eqnarray}
Comparing coefficients yields,
\begin{equation} (2 l +1)c_n / k_n = i^{-l} \sum_j e^{- i (\delta_l)_{n j}} (n_l)_{n j} (A_l)_j. \label{eq:comparecoeffs} \end{equation}
This step corresponds to imposing the conditions, from Eq. \ref{eq:scatteringsolution}, that (1) the scattered wave is purely outgoing, and (2) the normalization of the $n$th component of the initial plane wave is set by $c_n$.

Let us define an $N \times N$ matrix $(M_l)_{i j} = e^{- i (\delta_l)_{i j}} (n_l)_{i j}$, describing the asymptotics of the basis solutions for each $l$, and an $N \times N$ matrix $C_{n j} = (\vec{c_j})_n / k_n$, where $\vec{c_1}, \vec{c_2}, ..., \vec{c_N}$ are $N$-vectors that form a basis for possible incoming states (in the simplest example, we could simply choose $(c_j)_n = \delta_{jn}$). Let us also define an $N \times N$ matrix $A_l$ such that the $m$th column of $A_l$ is the coefficient vector $\vec{A}_l$ corresponding to the incoming state described by $\vec{c}_m$. Then Eq. \ref{eq:comparecoeffs} can be written in matrix form as $A_l = i^l (2 l + 1) M_l^{-1} C$. A basis of solutions to the Schr\"{o}dinger equation is then given by the columns of the matrix,
\begin{equation} \Psi = \sum_{l=0} P_l(\cos \theta) i^l (2 l + 1) R_l M_l^{-1} C. \end{equation}

As $r \rightarrow 0$, $R_l(r) \sim r^l$, so the lowest-$l$ partial wave will dominate (at least to the extent that the short-range interaction of interest can be approximated as a contact interaction). Writing $R_l(r) = \chi_l(r) / r$, the reduced Schr\"{o}dinger equation has the standard form,
\begin{equation} \frac{\hbar^2}{m_\chi} \chi_l''(r) = \left( \mathcal{V}(r) - m_\chi v^2 + \frac{\hbar^2}{m_\chi} \frac{l(l+1)}{r^2} \right) \chi_l(r), \label{eq:reducedSE} \end{equation}
and $R_l(r \rightarrow 0) = r^l \chi^{(l+1)}(0) / (l+1)!$. Consequently we can write,
\begin{equation} \Psi_{s \rm{-wave}}(r = 0) = \chi'(0) M_0^{-1} C, \quad \Psi_{l}(r \rightarrow 0) = P_l(\cos \theta) r^l \left( \frac{i^l (2 l + 1)}{(l+1)!} \right) \chi^{(l+1)}(0)M_l^{-1} C. \end{equation}
Multiplying the annihilation matrix by this matrix and its conjugate ($\Psi^{\dagger} \Gamma \Psi$) will give the Sommerfeld-enhanced annihilation cross-section. Specifically, the $m$th diagonal element of the resulting $N \times N$ matrix will give the enhancement for the ingoing state described by the vector $\vec{c}_m$ (since that element corresponds to $\vec{\Psi}_m^\dagger \Gamma \vec{\Psi}_m$). In the case of higher than $s$-wave modes, it no longer suffices to treat the annihilation interaction as point-like for determining the annihilation cross section, but the enhancement relative to the unperturbed annihilation cross section can be calculated in this simple framework by taking the limit as $r \rightarrow 0$.

Consider the unperturbed case with $\mathcal{V}(r) = \mathcal{V}(\infty)$, in which case $\mathcal{V}$ is purely diagonal. Then $\chi_l$ has an exact solution in terms of Bessel functions,
\[(\chi_l(r))_{i j} = \sqrt{r} \left( (C_1)_{i j} J_{l+1/2}(k_i r) + (C_2)_{i j} Y_{l + 1/2}(k_i r) \right). \]
Imposing the condition of regularity at the origin sets $C_2 = 0$, and the boundary condition of Eq. \ref{eq:largerbasis} sets $(C_1)_{i j} = \sqrt{\pi k_i / 2}  (n_l)_{i j}$, as well as fixing $\delta_l = 0$. Then as $r \rightarrow 0$ the unperturbed radial solution has the asymptotic form $(\chi_l(r \rightarrow 0))_{i j} = (n_l)_{i j} \sqrt{r} \sqrt{\pi k_i / 2} (k_i r/2)^{l + 1/2} / \Gamma(l + 3/2) = (n_l)_{i j} k_i^{l + 1} r^{l+1} / (2 l + 1)!!$, and $M_l = n_l$, which in turn yields $ (\Psi_{l}(r \rightarrow 0))_{i j} = P_l(\cos \theta) r^l \left(i^l (2 l + 1)/(l+1)! \right) \left((l+1)! / (2 l + 1)!! \right) k_i^l (\vec{c_j})_i $. 

Now comparing the perturbed case to the unperturbed result, the Sommerfeld enhancement where the $l$th partial wave dominates, for the ingoing state described by $\vec{c_m}$, becomes,
\begin{equation} S_m = \left(\frac{(2 l + 1)!!}{(l + 1)!} \right)^2 \frac{\left\{(\chi^{(l+1)}(0)M_l^{-1} C)^\dagger \cdot \Gamma \cdot (\chi^{(l+1)}(0)M_l^{-1} C) \right\}_{mm}}{\sum_{i, j} (k_i k_j)^l (\vec{c_m}^*)_i \Gamma_{i j} (\vec{c_m})_j}. \end{equation}
In the case where $N = 1$, so $\vec{c} \rightarrow 1$, $C \rightarrow 1/k$ and $M_l \rightarrow e^{-i \delta_l}$, this expression reduces to,
\begin{equation} S = \left| \frac{(2 l + 1)!! \chi^{(l+1)}(0)}{(l + 1)! k^{l+1}} \right|^2, \end{equation}
in agreement with \cite{Iengo:2009ni, Cassel:2009wt}.

Using this result requires us to know the phase shifts described by $M_l$, since while the enhancement factor is real by definition, there are several different phase shifts involved and $\Psi^\dagger \Gamma \Psi$ contains cross terms (in contrast to the $N = 1$ case where the phases cancel trivially, and so can be set to any arbitrary value). The phase shifts can be determined by solving the Schr\"{o}dinger equation for an appropriate basis of regular solutions. 

An alternate form for the enhancement can be obtained in terms of irregular solutions to the (reduced) Schr\"{o}dinger equation (Eq. \ref{eq:reducedSE}), rather than the regular solution $\chi(r)$. Consider the matrix $\rho_l(r)$ satisfying the reduced Schr\"{o}dinger equation, such that $(\rho_l)_{ij}(r) \rightarrow T_{ij} e^{i k_i r}$ as $r \rightarrow \infty$ (i.e. a purely outgoing spherical wave), for $T$ some constant (complex) matrix to be determined, and $(\rho_l)_{ij}(r \rightarrow 0) \rightarrow r^{-l}$ (this is the correct asymptotic form for the $l$th partial wave at small $r$, provided the potential grows more slowly than $1/r^2$ at small $r$, so the $l(l+1)/r^2$ term dominates). Note that if $k_i$ is imaginary, the ``outgoing wave'' boundary condition is replaced by requiring that component of the wavefunction to be exponentially decaying as $r \rightarrow \infty$.

Now consider the quantity $W_l = \rho_l^T \chi_l' - \rho_l'^T \chi_l$. It is easy to check from the Schr\"{o}dinger equation that $W_l$ is independent of $r$ so long as the potential matrix $\mathcal{V}$ is symmetric. But writing $\chi_l(r \rightarrow 0) = r^{l+1} \chi^{(l+1)}(0) / (l+1)!$, we have,
\begin{equation}W_l(r=0) = \frac{2 l+1}{(l+1)!} \chi^{(l+1)}(0), \end{equation}
\begin{eqnarray}W_l(r \rightarrow \infty)_{i j} & = & \sum_m \left( T_{m i} e^{i k_m r} k_m \cos \left(k_m r  - \frac{1}{2} l \pi + (\delta_l)_{m j} \right) n_{m j} \right. \nonumber \\ & & \left. - i k_m T_{m i} e^{i k_m r} \sin \left(k_m r  - \frac{1}{2} l \pi + (\delta_l)_{m j} \right) n_{m j} \right) \nonumber \\  & = & i^l \sum_m T_{m i} e^{- i (\delta_l)_{m j}} n_{m j} k_m \nonumber \\ & = & i^l \sum_m T_{m i} (M_l)_{m j} k_m.  \end{eqnarray}

Taking $C$ to be the diagonal matrix with elements $1/k_n$ (i.e. the incoming states $c_n$ are the natural basis states), $C^{-1}$ becomes the diagonal matrix with elements $k_n$, i.e. $(C^{-1})_{a b} = k_a \delta_{a b}$. Then we can write,
\begin{equation} W_l(r \rightarrow \infty)_{i j} = i^l \sum_m T_{m i} (M_l)_{a j} (C^{-1})_{m a} = i^l (T^T C^{-1} M_l)_{i j}, \end{equation} 
and by the $r$-independence of $W_l$ we have,
\begin{equation} T^T C^{-1} M_l = i^{-l} \frac{2 l+1}{(l+1)!} \chi^{(l+1)}(0)  \quad \Rightarrow \quad \chi^{(l+1)}(0) M_l^{-1} C = i^l \frac{(l+1)!}{2 l+1} T^T. \end{equation}
The enhancement to annihilation for particles initially in the $m$th state, if the $l$th partial wave dominates, can then be written in the form,
\begin{equation} S_m = \left(\frac{(2 l - 1)!!}{k_m^l} \right)^2 \frac{\left(T^* \Gamma T^T \right)_{mm}}{\Gamma_{m m}}. \label{eq:sommerfeldexpr}\end{equation}
In the $s$-wave case and assuming $\Gamma$ real and symmetric, we recover the usual expression for the Sommerfeld-enhanced annihilation cross-section in an $N$-state system (see e.g. \cite{Cirelli:2007xd}). In the absence of any potential, we recover $S=1$ as required (note the identity $\Gamma(l+1/2) = (2 l - 1)!! \sqrt{\pi}/2^l$).
 
\section{The enhancement in a two-state system with off-diagonal interaction}
\label{sec:parametrics}

Consider the case of a U(1) vector interaction coupling two states with some small mass splitting $\delta$, where the interacting particles cannot carry conserved charge and thus the long-range interaction between the $|1 \rangle$ and $|2 \rangle$ states is purely off-diagonal. Then the $4 \times 4$ potential matrix coupling the 2-particle states ($|11 \rangle, \, |1 2 \rangle, \, |2 1 \rangle, \, |2 2 \rangle$) is block diagonal (the $|11 \rangle$ and $|22 \rangle$ states only couple to each other, and the same for the $|1 2\rangle$ and $|2 1 \rangle$ states). We will be interested in the Sommerfeld enhancement for a pair of particles initially in the ground state, so we wish to solve the Schr\"{o}dinger equation with the $2 \times 2$ matrix potential which couples the $|11 \rangle$ and $|22 \rangle$ states, given by \cite{ArkaniHamed:2008qn},
\begin{equation} \mathcal{V} = \left( \begin{array}{cc} 0 & - \hbar c \alpha \frac{e^{-m_\phi c r /\hbar}}{r} \\ - \hbar c \alpha\frac{e^{-m_\phi c r /\hbar}}{r}  &  2 \delta c^2 \end{array} \right), \end{equation}
where the first row corresponds to the ground state $|11 \rangle$. 
 
Define the dimensionless parameters $\epsilon_v = (v/c)/\alpha$, $\epsilon_\delta = \sqrt{2 \delta / m_\chi}/\alpha$, $\epsilon_\phi = (m_\phi/m_\chi)/\alpha$. Then rescaling the $r$ coordinate by $\alpha m_\chi c / \hbar$, we can rewrite the Schr\"{o}dinger equation as,
\begin{equation} \psi''(r) = \left( \begin{array}{cc} \frac{l(l+1)}{r^2}- \epsilon_v^2 & -\frac{e^{-\epsilon_\phi r}}{r} \\- \frac{e^{-\epsilon_\phi r}}{r}  &  \frac{l(l+1)}{r^2} + \epsilon_\delta^2 - \epsilon_v^2 \end{array} \right) \psi(r). \label{eq:dimensionless} \end{equation}

More generally, consider the matrix $ M = \left( \begin{array}{cc} \frac{l(l+1)}{r^2} - \epsilon_v^2  & - V(r) \\ - V(r)  & \frac{l(l+1)}{r^2} + \epsilon_\delta^2 - \epsilon_v^2  \end{array} \right) $
for some arbitrary scalar potential $V(r)$. The eigenvalues and eigenvectors of $M$ are given by,
\begin{equation} \lambda_\pm = -\epsilon_v^2 + \frac{l(l+1)}{r^2} + \frac{\epsilon_\delta^2}{2} \pm \sqrt{\left(\epsilon_\delta^2/2\right)^2 + V(r)^2}, \quad \psi_\pm = \frac{1}{\sqrt{2}} \left(  \begin{array}{c} \mp \sqrt{1 \mp \frac{1}{\sqrt{1 + (V(r)/ (\epsilon_\delta^2/2))^2}}}  \\ \sqrt{1 \pm \frac{1}{\sqrt{1 + (V(r)/ (\epsilon_\delta^2/2))^2}}} \end{array} \right). \label{eq:diag} \end{equation} 

There is a transition in the behavior of the eigenvalues and eigenvectors at $V(r) \sim \epsilon_\delta^2/2$. For $V(r) \gg \epsilon_\delta^2/2$, a 45$^\circ$ rotation approximately diagonalizes $M$, with the eigenvalues and eigenvectors taking the form,
\begin{equation} \lambda_\pm \approx \frac{l(l+1)}{r^2} -\epsilon_v^2 + \frac{\epsilon_\delta^2}{2} \pm V(r), \quad \psi_\pm \approx \frac{1}{\sqrt{2}} \left(  \begin{array}{c} \mp 1  \\ 1 \end{array} \right). \end{equation} 
For $V(r) \ll \epsilon_\delta^2/2$, on the other hand, $M$ is already approximately diagonal, with eigenvalues and eigenvectors:
\begin{equation} \lambda_+ \approx \frac{l(l+1)}{r^2} -\epsilon_v^2 + \epsilon_\delta^2 + \frac{V(r)^2}{\epsilon_\delta^2},  \, \lambda_- \approx \frac{l(l+1)}{r^2} -\epsilon_v^2 - \frac{V(r)^2}{\epsilon_\delta^2}, \quad \psi_+ \approx \left(  \begin{array}{c} 0  \\ 1 \end{array} \right), \, \psi_- \approx \left(  \begin{array}{c} 1  \\ 0 \end{array} \right). \label{eq:largerdiag} \end{equation} 
In these regimes, because there is an $r$-independent diagonalization of the Schr\"{o}dinger equation, if $\phi_\pm''(r) = \lambda_\pm(r) \phi_\pm(r)$, then $\phi_\pm(r) \psi_\pm$ is an approximate solution to the matrix ODE $\psi''(r) = M \psi(r)$. In the intermediate transition regime, however, where $V(r) \sim \epsilon_\delta^2/2$, the eigenvectors are clearly not constant with respect to $r$, and attempting to diagonalize the differential equation gives rise to additional (non-diagonal) terms depending on the derivatives of the diagonalizing matrix. We can regard the behavior of the solution in this intermediate region as a nonadiabatic transition between the eigenstates, of Rosen-Zener-Demkov type (a general class of nonadiabatic transitions related to models first studied by Rosen and Zener \cite{PhysRev.40.502} and Demkov \cite{Demkov}). We define the transition radius $r_T$ by $V(r_T) = \epsilon_\delta^2/2$.

\subsection{The limit of zero mass splitting}

In the limit as $\epsilon_\delta \rightarrow 0$, $r_T \rightarrow \infty$, and the potential matrix is always exactly diagonalizable by a 45$^\circ$ rotation: the Sommerfeld enhancement then follows directly from the enhancements (or suppressions) for the attractive and repulsive scalar potentials. In particular, if all elements of the annihilation matrix $\Gamma$ are identical, then the enhancement to $s$-wave annihilation for particles originating in the ground state is given by,
\begin{eqnarray} S_{11} & = & \left( \frac{1}{2} \left( \begin{array}{c|c} \phi_+(\infty) + \phi_-(\infty) & \phi_-(\infty) - \phi_+(\infty)  \\ \hline \phi_-(\infty) - \phi_+(\infty)   & \phi_+(\infty) + \phi_-(\infty) \end{array} \right)^* \left( \begin{array}{cc} 1 & 1 \\ 1 & 1 \end{array} \right) \frac{1}{2} \left( \begin{array}{c|c} \phi_+(\infty) + \phi_-(\infty) & \phi_-(\infty) - \phi_+(\infty)  \\ \hline \phi_-(\infty) - \phi_+(\infty)   & \phi_+(\infty) + \phi_-(\infty) \end{array} \right)^T \right)_{11} \nonumber \\ & = & |\phi_-(\infty)|^2, \end{eqnarray}
where $\phi_-$ and $\phi_+$ are, respectively, the solutions to the scalar Schr\"{o}dinger equation with attractive and repulsive potentials, with $\phi_-(0) = \phi_+(0) = 1$ (for the case of $l > 0$ the same argument holds, with the appropriate rescaling of the enhancement (Eq. \ref{eq:sommerfeldexpr}) and redefinition of $\phi_\pm(0)$). So in this limit we precisely recover the usual result for the enhancement, and the repulsive component does not contribute.

\subsection{Regimes where the enhancement is negligible}

Before deriving a detailed expression for the enhancement, it is useful to consider the limits where the enhancement must approach unity, in terms of the dimensionless parameters $\epsilon_v$, $\epsilon_\phi$ and $\epsilon_\delta$.

The characteristic Bohr energy of the Coulomb potential is given by $\sim \alpha^2 m_\chi c^2$. The presence of a Yukawa cutoff can only reduce this energy. If this scale is small compared to the kinetic energy, the potential is irrelevant and the enhancement must be $\mathcal{O}(1)$: this condition is equivalent to $m_\chi v^2 \gtrsim \alpha^2 m_\chi c^2 \Leftrightarrow \epsilon_v \gtrsim 1$. If the range of the Yukawa potential is much smaller than the Bohr radius, then again the potential cannot significantly affect the wavefunction. This criterion becomes $1/m_\phi \lesssim 1/\alpha m_\chi$, i.e. $\epsilon_\phi \gtrsim 1$. These conditions are familiar from the usual $\delta=0$ case (see e.g. \cite{ArkaniHamed:2008qn}).

We might suppose that if the energy of the mass splitting were significantly larger than the kinetic energy of the system, the higher state could not be excited and no interaction would occur. However, if the potential energy is sufficiently large, it can allow excitations at small $r$: the relevant criterion is therefore whether the energy of excitation $2 \delta c^2$ exceeds $\alpha^2 m_\chi c^2$. This condition is equivalent to $\epsilon_\delta \gtrsim 1$. Since this criterion is new to the case with $\delta > 0$, Fig. \ref{fig:deltacutoff} displays the cutoff in the enhancement at $\epsilon_\delta \gtrsim 1$ for sample $\epsilon_v, \, \epsilon_\phi$ parameters.

\FIGURE[h*]{
\includegraphics[width=0.35\textwidth]{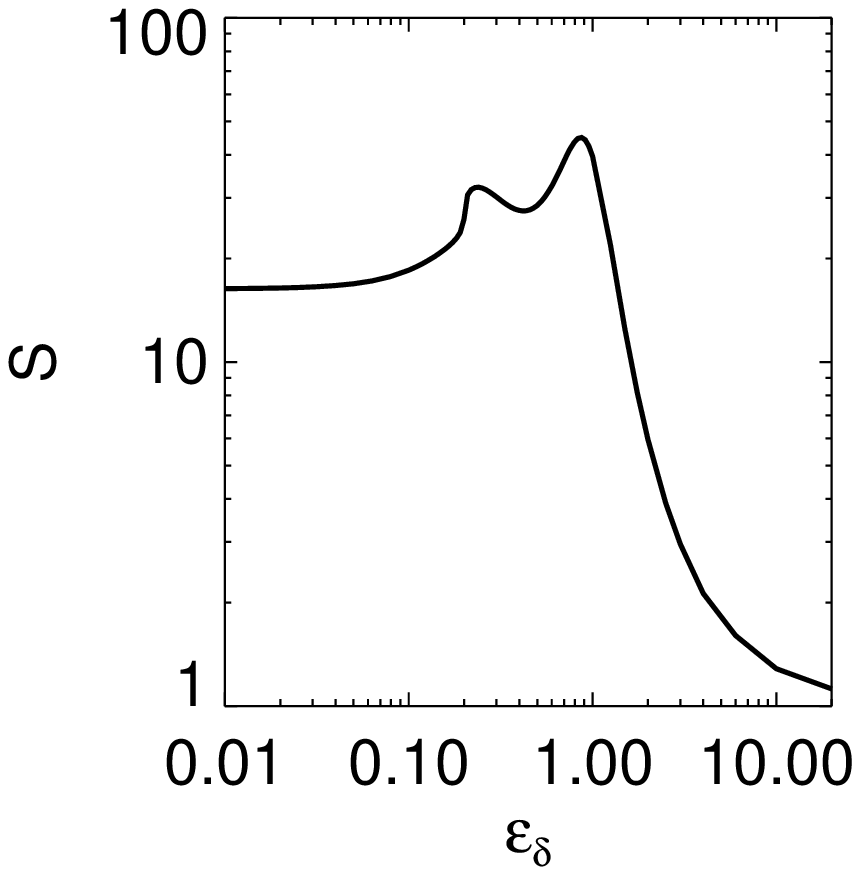}
\caption{\label{fig:deltacutoff} 
The enhancement $S$ as a function of $\epsilon_\delta$ for $\epsilon_v = 0.2$, $\epsilon_\phi = 0.1$, demonstrating the cutoff at large $\epsilon_\delta$. $S$ was computed numerically using \texttt{Mathematica}.}
}

For $l > 0$ it is possible for the $l(l+1)/r^2$ term to dominate the effective scalar potential everywhere, in which case the potential does not significantly modify the wavefunction and there is no significant Sommerfeld enhancement. This occurs for a Yukawa potential if $l(l+1) \gtrsim 1/\epsilon_\phi$, but in the presence of a finite mass splitting this can be true even for $\epsilon_\phi \rightarrow 0$. 

In the $\epsilon_\phi \rightarrow 0$ limit, if $l(l+1)/r_T^2 \gg \epsilon_\delta^2/2$, then since $l(l+1)/r^2$ grows faster than $V(r)$ as $r \rightarrow 0$, this term dominates for all $r \lesssim r_T$. For $r > r_T$, the effective scalar potential becomes $V(r)^2/\epsilon_\delta^2$, which for a Coulomb (or Yukawa) potential falls faster than $1/r^2$ as $r \rightarrow \infty$. Thus if $l(l+1) \gg (\epsilon_\delta r_T)^2/2$, the potential term is always subdominant; only partial waves with $l(l+1) \lesssim (\epsilon_\delta r_T)^2/2$ experience a large enhancement (note that since $r_T \le 2/\epsilon_\delta^2$, this condition requires that $l(l+1) \lesssim 2/\epsilon_\delta^2$). Similarly, if $\forall r$ either $l(l+1)/r^2 \gg V(r)$ or $\epsilon_v^2 \gg V(r)$, the enhancement also cuts off. This is a more stringent constraint than the previous one when $\epsilon_v \gtrsim \epsilon_\delta$; if $V(r_V) = \epsilon_v^2$, then a significant enhancement requires $l(l+1) \lesssim (\epsilon_v r_V)^2$ (which in turn requires $l(l+1) \lesssim 1/\epsilon_v^2$).

More generally, if $\epsilon_\phi \ne 0$, the constraints above generalize to the requirement that for the $l$th partial wave to experience a significant Sommerfeld enhancement, the potential term must dominate the eigenvalue for at least some range of $r \lesssim r_T$: if the potential term is subdominant at $r \sim r_T$ then the effective scalar potential falls faster than the $l(l+1)/r^2$ term at larger $r$.

In summary, a significant enhancement requires
\[ \epsilon_v, \, \epsilon_\delta, \, \epsilon_\phi \lesssim 1,\]
and only partial waves with $l(l+1) \lesssim 1/\epsilon_\phi$, and satisfying the condition $V(r) \gtrsim \max(|\epsilon_v^2 - \epsilon_\delta^2/2|, l(l+1)/r^2)$ for some $r \lesssim r_T$ can be enhanced.

\[\]

\section{An approximate solution to the Schr\"{o}dinger equation}
\label{sec:approxsol}

To obtain an approximate solution to the matrix Schr\"{o}dinger equation in the form of Eq. \ref{eq:dimensionless}, we note that if the Yukawa potential $V(r)$ is replaced with an exponential potential $\tilde{V}(r) = V_0 e^{-\mu r}$, then Eq. \ref{eq:dimensionless} is exactly solvable in the $s$-wave case \cite{PhysRevA.49.265}. The relevant properties of this solution are described in Appendix \ref{sec:expexactsol}. For $r \gtrsim 1/\epsilon_\phi$, the Yukawa potential can be approximated well by an exponential potential with an appropriate choice of parameters; the resulting solution for the wavefunction will be described in detail in \S \ref{sec:exactsolsfinal}.

At $r \lesssim 1/\epsilon_\phi$, the Yukawa and exponential potentials behave very differently. However, if the approximation of $r$-independent diagonalization discussed in \S \ref{sec:parametrics} holds over this region -- i.e. if $r_T \gtrsim 1/\epsilon_\phi$ -- then solving the matrix Schr\"{o}dinger equation in this region reduces to solving a pair of scalar Schr\"{o}dinger equations with attractive and repulsive Yukawa potentials. This problem will be discussed in detail in \S \ref{sec:innerregionsol}: in summary, we will show that the WKB approximation is valid (except possibly for an isolated turning point, denoted $r=r^*$, in the repulsive case) for $r \gtrsim 1$, breaking down when $V(r) \lesssim \epsilon_\phi^2$ (which occurs at $r \sim 1/\epsilon_\phi$), or when the diagonalization approximation fails at $r \approx r_T$. For $r \lesssim 1$ the Yukawa potential can instead be well approximated by a Coulomb potential (since we have assumed $\epsilon_\phi \lesssim 1$) and the scalar Schr\"{o}dinger equation can be solved exactly.

We will define $r=r_M$ to be the radius at which the WKB wavefunction (valid for $1 \lesssim r \lesssim \mathrm{min}(1/\epsilon_\phi, r_T)$) is matched onto the solution obtained by approximating the Yukawa potential with an exponential potential (valid for $r \gtrsim 1/\epsilon_\phi$). In this work we will choose the matching radius $r_M$ such that $V(r_M) = \max(\epsilon_\delta^2/2, \epsilon_\phi^2)$: this ensures that the WKB and diagonalization approximations are both valid for $1 \lesssim r \lesssim r_M$ (except possibly at some isolated turning point). Fig. \ref{fig:matchpoints} summarizes the different regions and the approaches employed to obtain approximate wavefunctions in each regime.

\FIGURE[h]{
  \includegraphics[width=0.8\textwidth]{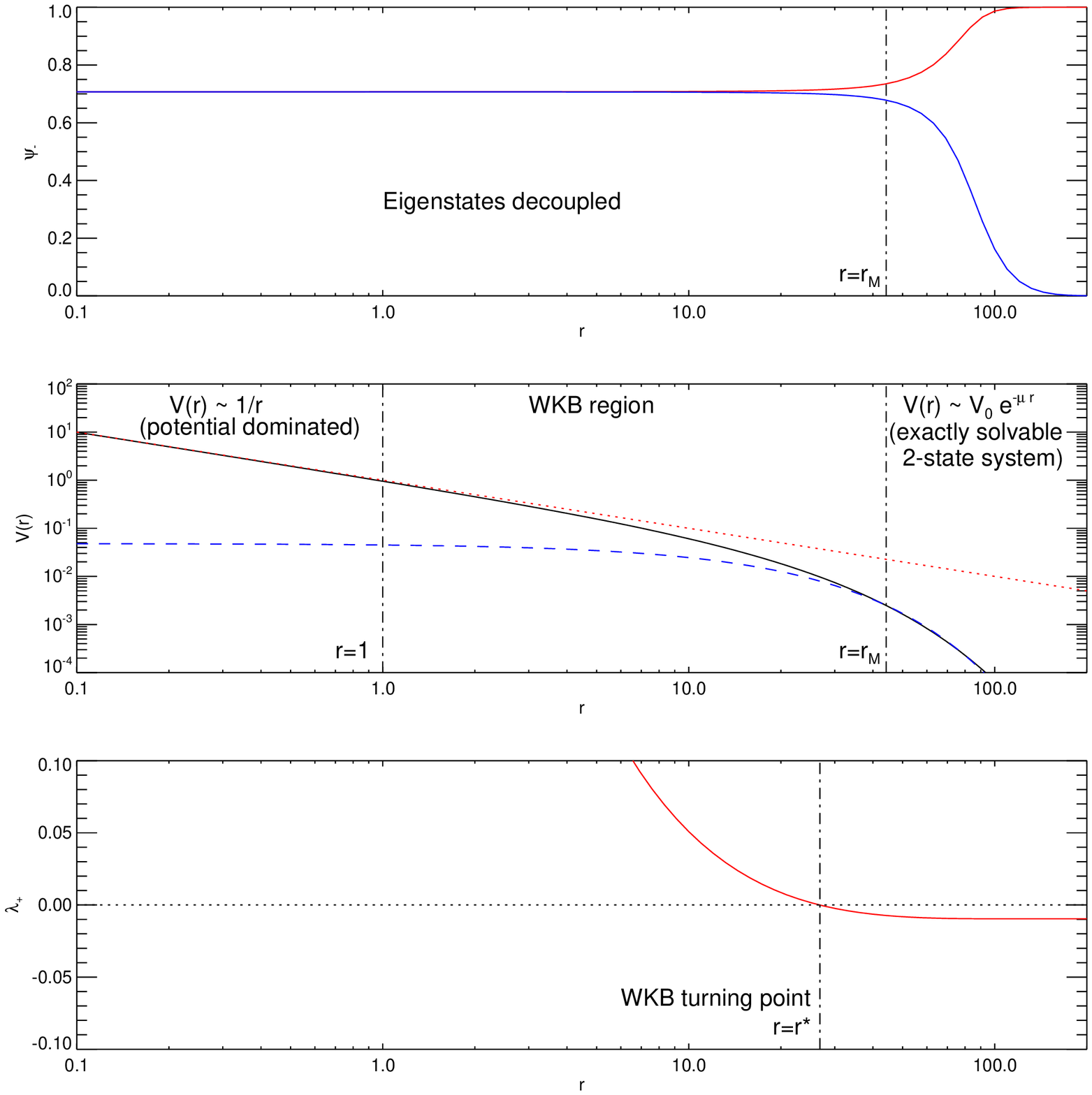}
\caption{\label{fig:matchpoints} 
Example of the different $r$-regimes and matching points $r=1, r^*, r_M$, for a sample parameter set ($\epsilon_v = 0.1, \, \epsilon_\delta = 0.02, \, \epsilon_\phi = 0.05$). \emph{Top panel:} The ground-state (\emph{red}) and excited-state (\emph{blue}) components of the eigenstate $\psi_-$ (Eq. \ref{eq:diag}), as a function of $r$. When the eigenstate components are nearly constant, as for $r \lesssim r_M$ ($\le r_T$), the eigenstates evolve independently and the problem reduces to solving two decoupled scalar Schr\"{o}dinger equations. $r \gtrsim r_M$ contains the non-adiabatic transition region discussed in \S \ref{sec:parametrics}. \emph{Middle panel:} The Yukawa potential (\emph{solid black line}) and the approximate potentials we employ, in their regimes of validity. In the $r \lesssim r_M$ region where the eigenstates are decoupled, for $r \lesssim 1$ the potential dominates and is well approximated by $V(r) \approx 1/r$ (\emph{red dotted line}), whereas for $1 \lesssim r \lesssim r_M$ the WKB approximation is employed to obtain an approximate wavefunction. At $r \gtrsim r_M$ the WKB approximation may break down, but there $V(r) \approx V_0 e^{-\mu r}$ (\emph{dashed blue line}). \emph{Bottom panel:} The eigenvalue $\lambda_+$ (\emph{red}), which passes through zero (signalling a breakdown in the WKB approximation) at $r=r^*$; the WKB wavefunction must be continued across $r=r^*$ as described in \S \ref{sec:innerregionsol}. }}

The criterion that $r_T \gtrsim 1/\epsilon_\phi$ is equivalent to requiring $W(2 \epsilon_\phi / \epsilon_\delta^2) \gtrsim 1$ (here $W$ denotes the Lambert $W$-function), which is true provided that $\epsilon_\phi \gtrsim \epsilon_\delta^2/2$. Provided that this condition holds, the non-adiabatic transition described in \S \ref{sec:parametrics} occurs close to the range of the interaction, where the Yukawa potential begins falling steeply. If this condition does not hold, then the Yukawa potential is still close to a $1/r$ potential when the non-adiabatic transition occurs, and is not well approximated by an exponential potential for all $r \gtrsim r_T$. Thus our results may be less accurate in this regime, although if the kinetic energy term dominates for all $r \gtrsim r_T$, which is true if $\epsilon_v^2 \gg \epsilon_\delta^2$, then the poor approximation of the true potential is irrelevant. For a Coulomb potential, a better approximate wavefunction might be obtained by a straightforward extension of this method: approximating the true potential by an exponential potential only for $r \sim r_T$, and matching onto a known solution for $r \gtrsim r_T$. However, the condition $\epsilon_\phi \gtrsim \epsilon_\delta^2/2$ corresponds to requiring $\delta \lesssim \alpha m_\phi$, which is generically satisfied for the phenomenologically motivated models of greatest interest \cite{ArkaniHamed:2008qn}, and in this part of parameter space our method is very accurate.

%\begin{table}
%\caption{Summary of regions of validity for approximate $s$-wave wavefunctions.}
%\label{tab:regions}
%\begin{tabular}{@{}lcccc}
%\hline
%Solution type  & Requirements & Regime of validity & Section no. \\
%\hline
%Small-$r$ exact solution for Coulomb potential & $V(r) \approx 1/r \gg \epsilon_\delta^2, \, \epsilon_v^2$ & $r \ll 1/\epsilon_\delta^2, \, 1/\epsilon_v^2, \, 1/\epsilon_\phi$ & \S \ref{sec:innerregionsol}            \\
%WKB solution for diagonalized Yukawa potential    & $V(r) \gtrsim \epsilon_\delta^2, \, \epsilon_\phi^2$, $r \gtrsim 1$         & $1 \lesssim r \lesssim r_M$             &  \S \ref{sec:innerregionsol}               \\
%Exact solution for exponential potential     & Determined by matching conditions      & $r \gtrsim r_M$             & \S \ref{sec:exactsolsfinal}                \\
%\hline
%\end{tabular}
%\end{table} 

\subsection{The small-$r$ solution}
\label{sec:innerregionsol}

For $V(r) \gg \epsilon_\delta^2/2$, as previously discussed we take the solution to the matrix Schr\"{o}dinger equation to be $\phi_+(r) \psi_+(r) + \phi_-(r) \psi_-(r)$, where $\psi_\pm(r)$ are the energy eigenstates and $\phi_\pm(r)$ are solutions to the scalar Schr\"{o}dinger equation $\phi''_\pm(r) = (l(l+1)/r^2 - \epsilon_v^2 + \epsilon_\delta^2/2 \pm V(r))\phi_\pm(r)$. Let us first specialize to the $s$-wave case with $l=0$.

We will employ the WKB approximation to find approximate wavefunctions in the small-$r$ region. When the constant term $-\epsilon_v^2 + \epsilon_\delta^2/2$ dominates, which is possible even if $V(r) \gtrsim \epsilon_\delta^2/2$ provided $\epsilon_v \gtrsim \epsilon_\delta$, the WKB approximation holds trivially. In the potential-dominated regime, $\lambda_\pm(r) \sim \pm V(r)$, the validity of the WKB approximation is determined by the parameter,
\[ |\sqrt{V(r)}'/V(r)| = (1/2)\sqrt{r} e^{\epsilon_\phi r /2} (\epsilon_\phi + 1/r).\]
For $r \ll 1/\epsilon_\phi$, the approximation is good so long as $1/2 \sqrt{r} \lesssim 1$, and breaks down at $r \sim 1$. At $r \sim 1/\epsilon_\phi$, where the potential begins falling rapidly, the condition for validity of the WKB approximation becomes $\epsilon_\phi / \sqrt{V(r)} \lesssim 1$, so the approximation breaks down when $V(r) \sim \epsilon_\phi^2$. By the assumption of a significant enhancement, $1/\epsilon_\phi \gtrsim 1$, so there is an intermediate range of $r$ where the WKB approximation is valid. At $r \lesssim 1$ we can approximate $V(r) \approx 1/r$ and assume the potential term dominates (since $\epsilon_v^2 \lesssim 1$ by assumption), solve the resulting ODE exactly, and match onto the WKB solution for $r \gtrsim 1$.

Note that we can now see explicitly that if $\epsilon_\phi \gtrsim 1$ the enhancement must cut off, since this corresponds to the potential falling off sharply at $r \lesssim 1$, and in the $r \sim 0-1$ region the magnitude of the wavefunction does not evolve significantly. Similarly, if $\epsilon_v \gtrsim 1$ then the kinetic term begins to dominate at $r \lesssim 1$, and the wavefunction is well approximated by a plane wave from that radius outward. If $\epsilon_\delta \gtrsim 1$ then $r_T \lesssim 1$, and at $r \gtrsim r_T$ the effective attractive scalar potential for the ground state is given by $\sim 1/r^2 \epsilon_\delta^2$ (Eq. \ref{eq:largerdiag}) (for the excited state the sign of the effective potential is reversed). While the potential term dominates (and $r \gtrsim r_T$), the approximate solutions to the Schr\"{o}dinger equation scale as $r^{1/2 \pm \sqrt{1/4 - 1/\epsilon_\delta^2}}$ (or $r^{1/2 \pm \sqrt{1/4 + 1/\epsilon_\delta^2}}$ for the excited state). The ground state wavefunction is therefore oscillatory provided $\epsilon_\delta < 2$, with the amplitude scaling as $1/\sqrt{r}$, but for $\epsilon_\delta > 2$, the physical solution becomes nearly constant (exactly constant in the limit of large $\epsilon_\delta$). Thus the amplitude of the wavefunction does not grow significantly for $r \gtrsim r_T$ in this case, and as stated previously, no large enhancement is possible for $\epsilon_\delta \gg 1$.

Returning to our calculation of the small-$r$ wavefunctions, the eigenvalue $\lambda_+$ is negative when the kinetic term dominates but positive when the potential term dominates; if $\epsilon_v \gtrsim \epsilon_\delta$, it may possess a zero in this small-$r$ region ($V(r) \gtrsim \epsilon_\delta^2/2$). In this case the WKB approximation for the ``repulsed'' solution has a turning point in the inner region, which we must take into account.

Let us first consider the potential-dominated region $r \lesssim 1$, where the WKB approximation breaks down and we can set $\lambda_\pm \approx \pm 1/r$ (since by assumption $\epsilon_v, \, \epsilon_\phi \lesssim 1$). The solutions to the scalar Schr\"{o}dinger equations are then given simply by,
\begin{equation} \phi_-(r) = A_- \sqrt{r} J_1(2\sqrt{r}) - \pi\phi_-(0) \sqrt{r} Y_1(2\sqrt{r}), \quad \phi_+(r) = A_+ \sqrt{r} I_1(2\sqrt{r}) + 2 \phi_+(0) \sqrt{r} K_1(2\sqrt{r}). \label{eq:scalarSElowr} \end{equation}
At $r \gtrsim 1$, using the large-$r$ asymptotics of the Bessel functions, these solutions match smoothly onto the WKB solutions, 
\begin{equation} \phi_+ = \frac{1}{\lambda_+^{1/4}} \left( \frac{A_+}{2 \sqrt{\pi}}e^{\int_0^r \sqrt{\lambda_+(r')} dr'} + \left( \sqrt{\pi} \phi_+(0) - \frac{i}{2 \sqrt{\pi}} A_+ \right) e^{- \int_0^r \sqrt{\lambda_+(r')} dr'} \right), \end{equation}
\begin{equation} \phi_- = \frac{1}{\lambda_-^{1/4}} \left(- \frac{1}{2 \sqrt{\pi}} (A_- + i \pi \phi_-(0)) e^{\int_0^r \sqrt{\lambda_-(r')} dr'} + \frac{i}{2 \sqrt{\pi}} \left(A_- - i \pi \phi_-(0) \right) e^{- \int_0^r \sqrt{\lambda_-(r')} dr'} \right). \end{equation}

Let $r^*$ be the radius such that $V(r^*) = \epsilon_v \sqrt{\epsilon_v^2 - \epsilon_\delta^2}$, so $\lambda_+(r^*) = 0$ (in the case $\epsilon_v > \epsilon_\delta$). Then linearizing the potential across the turning point and matching to the WKB solutions on either side, we obtain the WKB solution for $r > r^*$:
\begin{eqnarray} \phi_+ & = & \frac{1}{|\lambda_+|^{1/4}} \left\{ \left( -i \frac{A_+}{2 \sqrt{\pi}}  + \frac{1}{2} \left(\sqrt{\pi} \phi_+(0) - \frac{i}{2 \sqrt{\pi}} A_+ \right) e^{-2 \int^{r*}_0 \sqrt{\lambda_+} dr'} \right) e^{i \pi/4 + \int^{r}_0 \sqrt{\lambda_+(r')} dr'} \right. \nonumber \\ & & \left. + \left( i  \frac{A_+}{2 \sqrt{\pi}}  e^{2 \int^{r*}_0 \sqrt{\lambda_+} dr'} + \frac{1}{2} \left(\sqrt{\pi} \phi_+(0) - \frac{i}{2 \sqrt{\pi}} A_+ \right)  \right) e^{- (i \pi /4 + \int^{r}_0 \sqrt{\lambda_+(r')} dr')} \right\} \end{eqnarray}

We can check this result in the limit $\epsilon_\delta \rightarrow 0, \, \epsilon_\phi \rightarrow 0$: we should recover the usual expression for the Gamow-Sommerfeld suppression factor due to a repulsive Coulomb potential \cite{sommerfeld,gamow},
\begin{equation*} S = \frac{\pi}{\epsilon_v} \left( \frac{1}{e^{\pi / \epsilon_v} - 1} \right). \end{equation*} 
In this case $r^* = 1/\epsilon_v^2$, and $\int^{r*}_0 \sqrt{\lambda_+} dr' = \pi / 2 \epsilon_v $. Imposing the boundary conditions that $\phi_+(0) = 1$ and $\phi(r)$ is purely outgoing at $r \rightarrow \infty$, we find,
\[ A_+  = i \pi / \left(  e^{\pi / \epsilon_v} - \frac{1}{2}  \right) \Rightarrow |\phi_+(\infty)| = \sqrt{\frac{\pi}{\epsilon_v}} \frac{e^{\pi / 2 \epsilon_v}}{ e^{\pi / \epsilon_v} - \frac{1}{2}}, \]
(note that only the integral of $\sqrt{\lambda_+}$ from $0$ to $r^*$ contributes to the amplitude, as beyond $r^*$ the integrand is imaginary), which yields a  suppression factor,
\[ S = \frac{\pi}{\epsilon_v} \left( \frac{1}{e^{\pi / \epsilon_v} - 1 + (1/4)e^{-\pi / \epsilon_v}} \right). \]
Dropping the exponentially small term in the denominator (since we have assumed that $\epsilon_v \ll 1$ in the matching with the $r \lesssim 1$ solution), we obtain the correct expression.

For the higher partial waves which experience a significant enhancement, there must exist a region where $V(r) \approx 1/r$ and is large compared to both the $l(l+1)/r^2$ term and the $\epsilon_v^2$ term. For $l=0$, this region extends inward to $r=0$; for higher partial waves the $l(l+1)/r^2$ term will dominate for $r \lesssim l(l+1)$. In the range of $r$ where $V(r) \approx 1/r$ and either $V(r)$ or $l(l+1)/r^2$ is the dominant term, we can approximate the eigenvalues as $\lambda_\pm = l(l+1)/r^2 \pm V(r)$, and solve the Schr\"{o}dinger equation exactly. For larger $r$, but still in the small-$r$ regime $r \lesssim 1/\epsilon_\phi$, the $l(l+1)/r^2$ term is subdominant (since it falls faster than $V(r)$), and so we can ignore the effects of non-zero $l$, and match onto the $s$-wave WKB wavefunction. The discussion of the validity of the WKB approximation in the $s$-wave case is equally applicable to this case: the WKB approximation may break down for $r \lesssim l(l+1)$ where the $l(l+1)/r^2$ term dominates, but by the assumption of a significant enhancement, $1/\epsilon_\phi \gtrsim l(l+1)$, so again there is a range of $r$ where the WKB approximation is valid.

Thus the only required change to the $s$-wave calculation is to approximate the eigenvalues at low $r$ by $\lambda_\pm(r) = l(l+1)/r^2 \pm 1/r$, replacing Eq. \ref{eq:scalarSElowr} with,
\begin{eqnarray} \phi_-(r) & = & \frac{\Gamma(2(l+1))}{2} A_- \sqrt{r} J_{2l+1}(2\sqrt{r}) - \frac{\pi}{\Gamma(2 l + 1)} \hat{\phi}_-(0) \sqrt{r} Y_{2l+1}(2\sqrt{r}), \nonumber \\ \phi_+(r) & = & \frac{\Gamma(2(l+1))}{2} A_+ \sqrt{r} I_{2l+1}(2\sqrt{r}) + \frac{2}{\Gamma(2 l+1)} \hat{\phi}_+(0) \sqrt{r} K_{2l+1}(2\sqrt{r}). \end{eqnarray}
Here $\hat{\phi}_+(0)$, $\hat{\phi}_-(0)$ are the coefficients of the $r^{-l}$ terms in $\phi_+(r)$, $\phi_-(r)$ respectively, as $r \rightarrow 0$. The large-$r$ limits of these solutions match onto the WKB wavefunctions derived for the $s$-wave case, with the replacements $\phi_\pm(0) \rightarrow \hat{\phi}_\pm(0) / \Gamma(2 l + 1)$, $A_\pm \rightarrow A_\pm \Gamma(2 l + 2)/2$. The WKB wavefunctions also gain an overall multiplicative factor of $(-1)^l$ (note that the $e^{\pm \int \sqrt{\lambda} dr}$ terms in the WKB wavefunctions should still be taken to refer to the $s$-wave eigenvalues $\lambda_\pm$).

\subsection{The large-$r$ solution}
\label{sec:exactsolsfinal}

Where the WKB approximation breaks down (at $V(r) \sim \epsilon_\phi^2$) and/or the diagonalization approximation fails (at $V(r) \sim \epsilon_\delta^2/2$), we match the previously derived wavefunctions onto the exact wavefunction in the presence of an off-diagonal exponential potential $\tilde{V}(r) = V_0 e^{-\mu r}$, outlined in Appendix \ref{sec:expexactsol}. For convenience, we define a new parameter $z = V_0^2 e^{-2 \mu r}/16 \mu^4$.

The parameters of the exponential potential, $V_0$ and $\mu$, are set by the requirement that the exponential potential should mimic the Yukawa potential for $r \gtrsim r_M$, the matching radius. We follow \cite{Cassel:2009wt} and require that $\int_{r_M}^{\infty} r V(r) dr = \int_{r_M}^\infty r \tilde{V}(r) dr$ \footnote{This criterion comes from solving the Lippman-Schwinger form of the Schr\"{o}dinger equation, to obtain $R_l(r)$ as a series in the potential $V$; in the limit $\epsilon_\delta, \epsilon_v \rightarrow 0$, the condition that $R_l(0)$ is correct to 1st order in the coupling $\alpha$ fixes $\int r V dr$ \cite{Cassel:2009wt}.}. We also impose the condition that $V(r_M) = \tilde{V}(r_M)$.

The parameter $\mu$ is then given by,
\begin{equation} \frac{1}{\epsilon_\phi} = \frac{1}{\mu} \left(r_M + \frac{1}{\mu}\right) \Rightarrow \mu = \epsilon_\phi \left( \frac{1}{2} + \frac{1}{2} \sqrt{1 + \frac{4}{\epsilon_\phi r_M}} \right). \label{eq:mu} \end{equation}
A more sophisticated prescription for the potential matching might yield better agreement; however, we will show that this simple prescription already gives good agreement with numerical results. 

We will employ Eq. \ref{eq:sommerfeldexpr} to determine the Sommerfeld enhancement; thus, we are interested in wavefunctions which are purely outgoing (or exponentially decaying) as $r \rightarrow \infty$. As discussed in Appendix \ref{sec:expexactsol}, this condition implies $C_1 = C_3 = 0$ in Eq. \ref{eq:exactsolapp}. The remaining coefficients $C_2, \, C_4$ are determined by matching to the solutions in the inner region. For ease in reading off the Sommerfeld enhancement, we write the exact solution for the exponential potential in the form,
\begin{eqnarray}\psi_1 & = & \eta e^{i \epsilon_v r} \,_0F_3\left[\{\},\left\{1-\frac{i \epsilon_v}{\mu},\frac{1}{2}-\frac{i \epsilon_v}{2 \mu}-\frac{i \sqrt{-\epsilon_\delta^2+\epsilon_v^2}}{2 \mu},\frac{1}{2}-\frac{i \epsilon_v}{2 \mu}+\frac{i \sqrt{-\epsilon_\delta^2+\epsilon_v^2}}{2 \mu}\right\},z \right] \nonumber \\ 
& & - \frac{\sqrt{z} \zeta e^{i \sqrt{\epsilon_v^2 - \epsilon_\delta^2} r}\,_0F_3\left[\{\},\left\{\frac{3}{2}-\frac{i \epsilon_v}{2 \mu}-\frac{i \sqrt{-\epsilon_\delta^2+\epsilon_v^2}}{2 \mu},\frac{3}{2}+\frac{i \epsilon_v}{2 \mu}-\frac{i \sqrt{-\epsilon_\delta^2+\epsilon_v^2}}{2 \mu},1-\frac{i \sqrt{-\epsilon_\delta^2+\epsilon_v^2}}{\mu}\right\},z\right]}{\frac{\epsilon_v^2}{4 \mu^2} + \left(\frac{1}{2} - i \frac{\sqrt{\epsilon_v^2 - \epsilon_\delta^2}}{2 \mu} \right)^2}   , \nonumber \end{eqnarray}
\begin{eqnarray}\psi_2 & = & -   \frac{\sqrt{z} \eta e^{i \epsilon_v r}\,_0F_3\left[\{\},\left\{\frac{3}{2}-\frac{i \epsilon_v}{2 \mu}-\frac{i \sqrt{-\epsilon_\delta^2+\epsilon_v^2}}{2 \mu},\frac{3}{2}-\frac{i \epsilon_v}{2 \mu}+\frac{i \sqrt{-\epsilon_\delta^2+\epsilon_v^2}}{2 \mu},1-\frac{i \epsilon_v}{\mu}\right\},z\right]}{(1/2 - i \epsilon_v / 2 \mu)^2 + (\epsilon_v^2 - \epsilon_\delta^2)/4 \mu^2}  \nonumber \\
& & + \zeta e^{i \sqrt{\epsilon_v^2 - \epsilon_\delta^2} r} \,_0F_3\left[\{\},\left\{1-\frac{i \sqrt{\epsilon_v^2 - \epsilon_\delta^2}}{ \mu},\frac{1}{2}-\frac{i \epsilon_v}{2 \mu}-\frac{i \sqrt{-\epsilon_\delta^2+\epsilon_v^2}}{2 \mu},\frac{1}{2}+\frac{i \epsilon_v}{2 \mu}-\frac{i \sqrt{-\epsilon_\delta^2+\epsilon_v^2}}{2 \mu}\right\},z\right]. \nonumber \end{eqnarray}
Here the $\,_p F_q$'s are hypergeometric functions. The elements of the Sommerfeld enhancement matrix $T$ relevant to particles initially in the ground state are then given by $\eta$ (calculated for the two LI boundary conditions at $r=0$, $\psi(0) = (0,1)$ and $\psi(0) = (1,0)$), and for particles initially in the excited state by $\zeta$. For $\epsilon_v < \epsilon_\delta$, only the $\eta$ coefficient is relevant, as the two-body excited state cannot be on-shell.

The matching procedure in the inner part of the transition region is described in Appendix \ref{sec:innermatch}. We define:
\begin{itemize}
\item $\tilde{\lambda}_\pm = -\epsilon_v^2 + \epsilon_\delta^2/2 \pm \sqrt{(\epsilon_\delta^2/2)^2 + \tilde{V}^2}$,
\item $r^\dagger$:  the radius (if any) at which the eigenvalue for the repulsed eigenstate of the exponential potential passes through zero, defined by $\tilde{V}(r^*) = \epsilon_v \sqrt{\epsilon_v^2 - \epsilon_\delta^2}$,
\item $r_S$: a (possibly negative) quantity chosen such that $V_0 e^{-\mu r_S} \gg \epsilon_v^2, \epsilon_\delta^2$. The corresponding value of $z$ is denoted $z_S = z(r_S)$.
\item $\epsilon_\Delta = \sqrt{|\epsilon_\delta^2 - \epsilon_v^2|}$.
\end{itemize}

We solve for $\eta, \, \zeta$ for arbitrary boundary conditions at the origin, obtaining for $s$-wave annihilation:

Below threshold:
\begin{equation} \eta = \frac{- 2 \sqrt{2} \pi ^2 e^{\frac{\epsilon_v \pi }{\mu}} \left(i e^{\frac{i \epsilon_\Delta \pi }{\mu}+i \theta_-} \phi_-(0)-e^{\theta_+} \phi_+(0)+e^{\frac{2 i \epsilon_\Delta \pi }{\mu}+2 i \theta_-+\theta_+} \phi_+(0)\right) }{\sqrt{\mu} \left(e^{\frac{\epsilon_v \pi }{\mu}}+e^{\frac{i \epsilon_\Delta \pi }{\mu}}\right) \left(-1+e^{\frac{\epsilon_v \pi +i \epsilon_\Delta \pi +2 \mu i \theta_-}{\mu}}\right)  \Gamma\left[1-\frac{i \epsilon_v}{\mu}\right] \Gamma\left[-\frac{i \epsilon_v}{2 \mu} + \frac{\epsilon_\Delta}{2 \mu} + \frac{1}{2} \right] \Gamma\left[-\frac{i \epsilon_v}{2 \mu} - \frac{\epsilon_\Delta}{2 \mu} + \frac{1}{2} \right]}, \nonumber \end{equation}

Above threshold:
\begin{eqnarray} \eta & = & \frac{- 2 \sqrt{2} \pi ^2 e^{\frac{\epsilon_v \pi }{\mu}}}{\sqrt{\mu} \Gamma\left[1-\frac{i \epsilon_v}{\mu}\right] \Gamma\left[\frac{1}{2} - i \left( \frac{ \epsilon_v + \epsilon_\Delta}{2 \mu} \right) \right] \Gamma\left[\frac{1}{2} - i \left( \frac{\epsilon_v - \epsilon_\Delta}{2 \mu} \right)\right] \left(e^{\frac{\epsilon_v \pi }{\mu}}+e^{\frac{\epsilon_\Delta \pi }{\mu}}\right) \left(-1+e^{\frac{\epsilon_v \pi +\epsilon_\Delta \pi +2 \mu i \theta_-}{\mu}}\right)}  \nonumber \\ & & \times \left( i \phi_-(0) e^{\frac{\epsilon_\Delta \pi }{\mu}+i \theta_-} + \frac{4 \phi_+(0) e^{\theta_+} \left( -1  +  e^{\frac{2 \epsilon_\Delta \pi }{\mu}+ 2 i \theta_-} \right)}{2 -e^{2 (\theta_T - \theta_Y)}+2 e^{2 (\theta^*+\theta_T - \theta_Y)}+e^{-2 \theta^*}} \right),  \nonumber \end{eqnarray}

\begin{eqnarray} \zeta & = & \frac{-2 \sqrt{2}\pi ^2 e^{\frac{\epsilon_\Delta \pi }{\mu}}}{\sqrt{\mu} \Gamma\left[1-\frac{i \epsilon_\Delta}{\mu}\right] \Gamma\left[\frac{1}{2}- i \left(\frac{ \epsilon_v+ \epsilon_\Delta}{2 \mu}\right)\right] \Gamma\left[\frac{1}{2} - i \left(\frac{\epsilon_\Delta - \epsilon_v}{2 \mu} \right)\right] \left(e^{\frac{\epsilon_v \pi }{\mu}}+e^{\frac{\epsilon_\Delta \pi }{\mu}}\right) \left(-1+e^{\frac{\epsilon_v \pi +\epsilon_\Delta \pi +2 \mu i \theta_-}{\mu}}\right)} \nonumber 
\\ & & \times \left(i \phi_-(0) e^{\frac{\epsilon_v \pi }{\mu}+i \theta_-} + \frac{4 \phi_+(0) e^{\theta_+} \left(1 - e^{\frac{2 \epsilon_v \pi }{\mu}+2 i \theta_-}\right) }{2 -e^{2 (\theta_T - \theta_Y)}+2 e^{2 ( \theta^*+ \theta_T - \theta_Y)}+e^{-2 \theta^*} } \right) . \label{eq:sommerfeldsols} \end{eqnarray}

In both the above and below threshold cases, the phases are given by,
\[  i \theta_- = \int^{r_M}_{r_S} \sqrt{\tilde{\lambda_-}}dr - 4 i z_S^{1/4} - \int^{r_M}_0 \sqrt{\lambda_-} dr, \quad \theta_+ = \int^{r_M}_{r_S} \sqrt{\tilde{\lambda_+}} dr - 4 z_S^{1/4} - \int^{r_M}_0 \sqrt{\lambda_+} dr, \]
\begin{equation} \theta_T = \int^{r_M}_{r^\dagger} \sqrt{\tilde{\lambda_+}} dr, \quad \theta^* = \int^{r*}_0 \sqrt{\lambda_+} dr, \quad \theta_Y = \int^{r_M}_0 \sqrt{\lambda_+} dr. \end{equation}

If $l > 0$, then there are two cases of interest. First, suppose that the $l(l+1)/r^2$ term is subdominant for all $r \gtrsim r_M$. Since the potential term is rapidly falling for $r \gtrsim r_M$, this requires that $l(l+1)/r_M^2 \lesssim \epsilon_v^2$. Then the large-$r$ $s$-wave wavefunction described above can be matched onto the rescaled WKB solution outlined in \S \ref{sec:innerregionsol}, and the only change from the $s$-wave case is that the elements of the enhancement matrix $T$ must all be rescaled by a factor $(-1)^l/\Gamma(2 l + 1)$. The Sommerfeld enhancement for $l > 0$ is then identical to the $s$-wave case, up to an overall rescaling by  $\left(\frac{(2 l - 1)!!}{\Gamma(2 l + 1) \epsilon_v^l} \right)^2$ (using Eq. \ref{eq:sommerfeldexpr}).

If $\epsilon_v \lesssim \sqrt{l(l+1)}/r_M$, this argument no longer holds, as the $l(l+1)/r^2$ term cannot be neglected for $r \gtrsim r_M$. Since the main focus of our work is the $s$-wave enhancement, we will not treat this case here.

\section{Properties of the Sommerfeld enhancement in an inelastic model}
\label{sec:analysis}

The results derived above can be used to compute the Sommerfeld enhancement for any annihilation matrix $\Gamma$, and particles in any initial state. However, in most applications we will be interested only in particles initially in the ground state. Furthermore, in the simplest models, all elements of the annihilation matrix $\Gamma$ are identical.

In this case, the $s$-wave annihilation enhancement for particles initially in the ground state (Eq. \ref{eq:sommerfeldexpr}) reduces to,
\begin{equation} S = |T_{11} + T_{12}|^2 = |\psi_1(\infty)|^2 \quad \text{with $\psi(0) = (1,1)$}. \end{equation}
This case is therefore the closest 2-state analogue of the purely attractive Yukawa potential, as the boundary condition at the origin sets the repulsive component $\phi_+(0)$ to zero (and $\phi_-(0) = \sqrt{2}$). The expressions derived in the previous section simplify dramatically in this case, yielding,
\begin{equation} S = \frac{2 \pi }{\epsilon_v}\sinh\left(\frac{\epsilon_v \pi }{\mu}\right) \left\{ \begin{array}{cc} \frac{1}{\cosh\left(\epsilon_v \pi/\mu\right)-\cos\left(\sqrt{\epsilon_\delta^2-\epsilon_v^2} \pi /\mu+2 \theta_-\right)} & \quad \epsilon_v < \epsilon_\delta, \\ \\ \frac{\cosh\left(\left(\epsilon_v+\sqrt{-\epsilon_\delta^2+\epsilon_v^2}\right) \pi/2 \mu\right) \text{sech}\left(\left(\epsilon_v-\sqrt{-\epsilon_\delta^2+\epsilon_v^2}\right) \pi/2 \mu \right)}{ \cosh\left(\left(\epsilon_v+\sqrt{-\epsilon_\delta^2+\epsilon_v^2}\right) \pi /\mu\right)-\cos(2 \theta_-)} &\quad \epsilon_v > \epsilon_\delta. \end{array} \right. \label{eq:simple} \end{equation}

While this expression still contains a numerical integral for $\theta_-$, it is well-behaved and fast to compute \footnote{Note that the $\int^{r_M}_{r_S} \sqrt{|\tilde{\lambda}_-|} dr - 4 z_S^{1/4}$ contribution to $\theta_-$ may be best computed by changing variables in the integral to $z$, and rewriting the second term as an integral, giving, 
\[\int^{z_S}_{\mathrm{max}(\epsilon_\phi^4,\epsilon_\delta^4/4)/16 \mu^4} \frac{1}{2 \mu z} \left(\sqrt{\epsilon_v^2 - \frac{\epsilon_\delta^2}{2} + \sqrt{\left(\frac{\epsilon_\delta^2}{2} \right)^2 + 16 \mu^4 z}} - 2 \mu z^{1/4} \right) dz -  \left(\frac{ \mathrm{max}(16 \epsilon_\phi^4,4 \epsilon_\delta^4)}{\mu^4} \right)^{1/4}. \]
This prevents the integral from diverging at large $z$, so $z_S$ can be taken to $\infty$.}.

More generally, note that whenever $\phi_+(0)$ appears in Eq. \ref{eq:sommerfeldsols}, it is accompanied by a factor of $e^{\theta_+}$. When $\epsilon_v < \epsilon_\delta$, $\theta_+$ is purely real (since $\lambda_+$, $\tilde{\lambda}_+$ are both positive for any $r \in [0,r_M]$), and can be written in the form,
\begin{eqnarray} \theta_+ & = & \int^{\infty}_{\mathrm{max}(\epsilon_\phi^4,\epsilon_\delta^4/4)/16 \mu^4}  \frac{1}{2 \mu z} \left(\sqrt{-\epsilon_v^2 + \frac{\epsilon_\delta^2}{2} + \sqrt{\left( \frac{\epsilon_\delta^2}{2} \right)^2 + 16 \mu^4 z}} - 2 \mu z^{1/4} \right) dz - \left(\frac{ \mathrm{max}(16 \epsilon_\phi^4,4 \epsilon_\delta^4)}{\mu^4} \right)^{1/4} \nonumber \\ & & - \int^{r_M}_0 \sqrt{\lambda_+} dr. \label{eq:thetaplusintegral} \end{eqnarray}
Here $z_S$ has been taken to $\infty$. The integrand in the first term satisfies the inequality,
\begin{eqnarray} \frac{1}{2 \mu z} \left(\sqrt{-\epsilon_v^2 + \frac{\epsilon_\delta^2}{2} + \sqrt{\left( \frac{\epsilon_\delta^2}{2} \right)^2 + 16 \mu^4 z}} - 2 \mu z^{1/4} \right) & \le & \frac{1}{2 \mu z} \left(\sqrt{-\epsilon_v^2 + \epsilon_\delta^2 + 4 \mu^2 \sqrt{z}} - 2 \mu z^{1/4} \right) \nonumber \\  & \le &   z^{-5/4} \left(\frac{\epsilon_\delta^2 - \epsilon_v^2}{8 \mu^2}\right). \label{eq:thetaplusineq} \end{eqnarray}
Here we have used the relations $\sqrt{a + b} \le \sqrt{a} + \sqrt{b}$ and $\sqrt{1 + a} \le 1 + a/2$ for $a$, $b$ real and positive.

Since the last term in Eq. \ref{eq:thetaplusintegral} is strictly negative (the integral is always real and positive), integrating Eq. \ref{eq:thetaplusineq}  yields,
\begin{equation} \theta_+ \le \left(\frac{\epsilon_\delta^2 - \epsilon_v^2}{8 \mu^2}\right) \left(4 (\mathrm{max}(\epsilon_\phi^4,\epsilon_\delta^4/4)/16 \mu^4)^{-1/4} \right) - \left(\frac{ \mathrm{max}(16 \epsilon_\phi^4,4 \epsilon_\delta^4)}{\mu^4} \right)^{1/4} \le 0. \end{equation}

So in the case where $\epsilon_v < \epsilon_\delta$, $\theta_+$ is real and negative. When $\epsilon_v > \epsilon_\delta$, the real part of $\theta_+$ becomes,
\[ \Re{\theta_+} = \int^{r^\dagger}_{r_S} \sqrt{\tilde{\lambda_+}} dr - 4 z_S^{1/4} - \int^{r^*}_0 \sqrt{\lambda_+} dr. \]
In this case we can write,
\[ \sqrt{\tilde{\lambda}_+} = \sqrt{-\epsilon_v^2 + \frac{\epsilon_\delta^2}{2} + \sqrt{\left( \frac{\epsilon_\delta^2}{2} \right)^2 + \tilde{V}(r)^2}} \le \sqrt{-\epsilon_v^2 + \epsilon_\delta^2 + \tilde{V}(r)} \le \sqrt{\tilde{V}(r)}, \]
and so,
\[ \int^{r^\dagger}_{r_S} \sqrt{\tilde{\lambda_+}} dr \le \int^{r^\dagger}_{r_S} \sqrt{V_0 e^{-\mu r}} dr = -\frac{2 \sqrt{V_0}}{\mu} \left( e^{-\mu r^\dagger/2} - e^{-\mu r_S / 2} \right) \le \frac{2 \sqrt{V_0}}{\mu} e^{-\mu r_S / 2} = 4 z_S^{1/4}. \]

Thus the real part of $\theta_+$ is always negative, and the contribution of the repulsed component is exponentially suppressed. To a first approximation, then, the elements of the enhancement matrix $T$ are determined solely by the coefficient of $\phi_-(0)$. Since both the $\psi(0) = (0,1)$ and $\psi(0) = (1,0)$ boundary conditions correspond to $\phi_-(0) = 1/\sqrt{2}$, the $s$-wave enhancement in this case (for particles initially in the ground state) becomes,
\[ S = S_\mathrm{att} (\Gamma_{11} + \Gamma_{12} + \Gamma_{21} + \Gamma_{22})/4 \Gamma_{11}, \]
where $S_\mathrm{att}$ is given by Eq. \ref{eq:simple}.

\subsection{Numerical tests of the approximation}

In Fig. \ref{fig:numchecks}, we compare $S$ computed using Eq. \ref{eq:simple} to the exact numerical solutions for the enhancement (computed using \texttt{Mathematica}), for $\epsilon_\delta = 0, \, 10^{-2} \, 10^{-1}$, in the regions of parameter space where Eq. \ref{eq:simple} is expected to hold and the numerical solution is stable. In all cases the approximate form correctly reproduces the numeric results to good accuracy, although it becomes less reliable as $\epsilon_\phi, \, \epsilon_v \rightarrow 1$, as expected.

\FIGURE[h]{
  \includegraphics[width=0.32\textwidth]{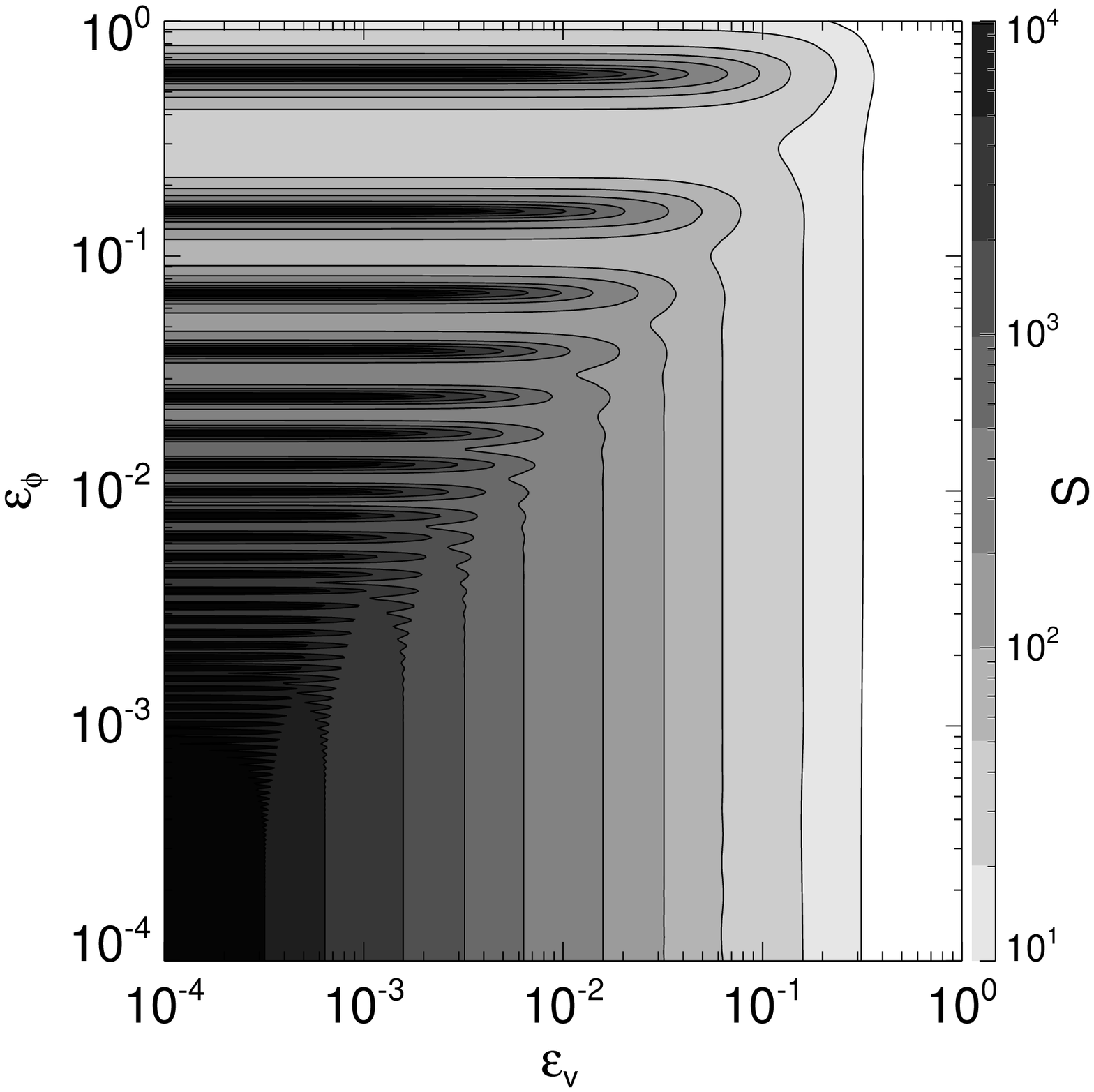}  
  \includegraphics[width=0.32\textwidth]{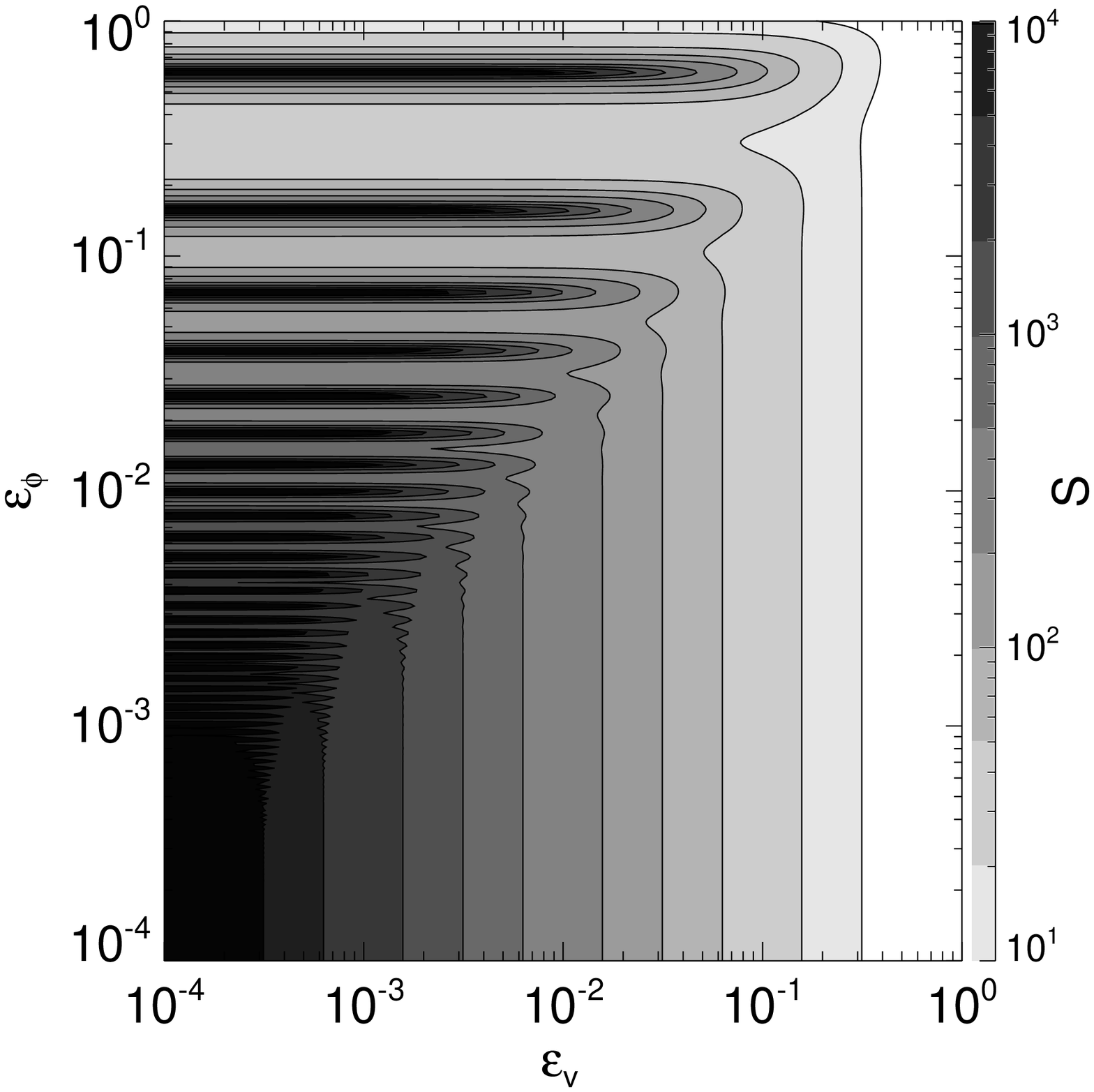}
  \includegraphics[width=0.32\textwidth]{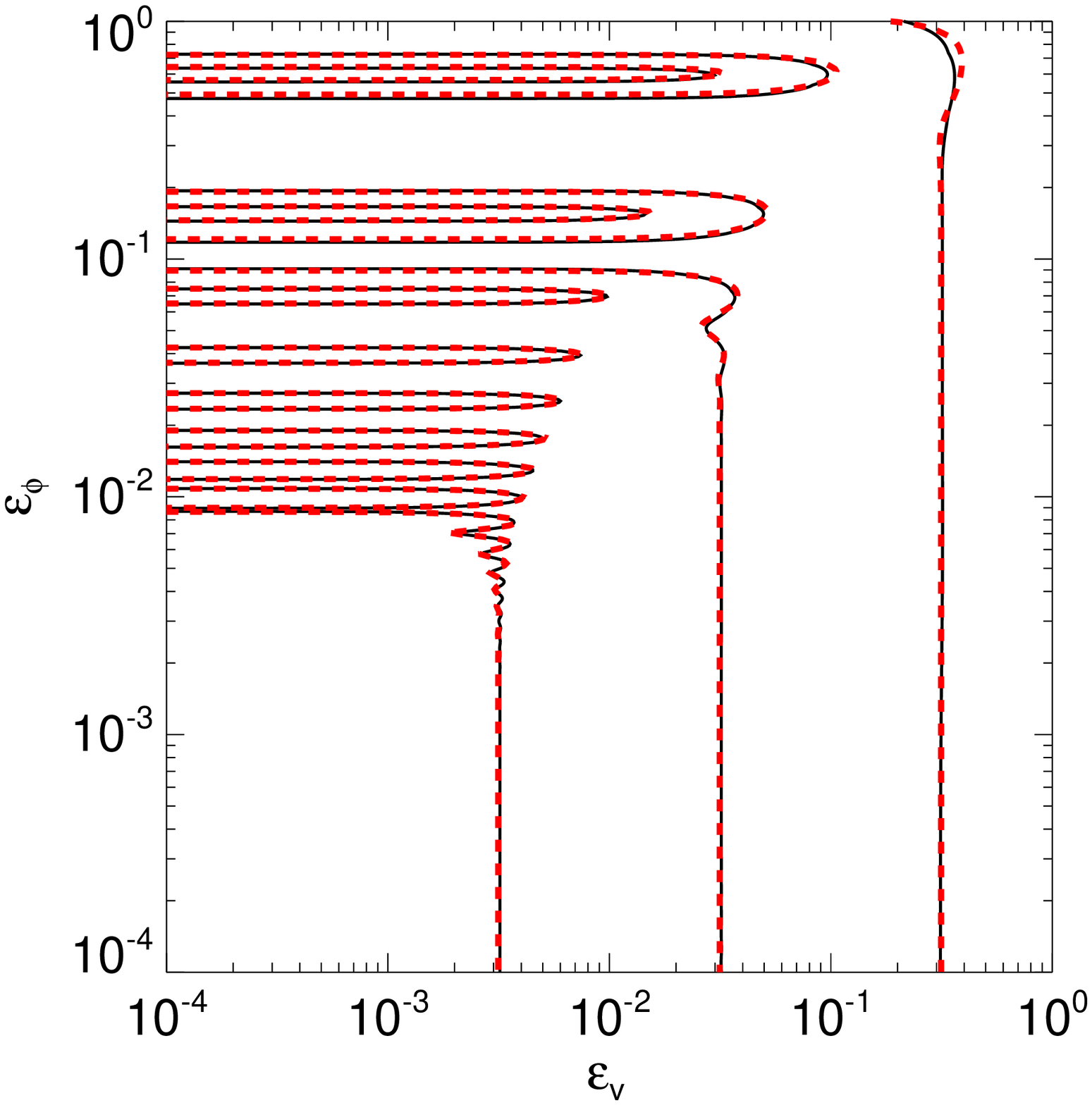} \\ 
  \includegraphics[width=0.32\textwidth]{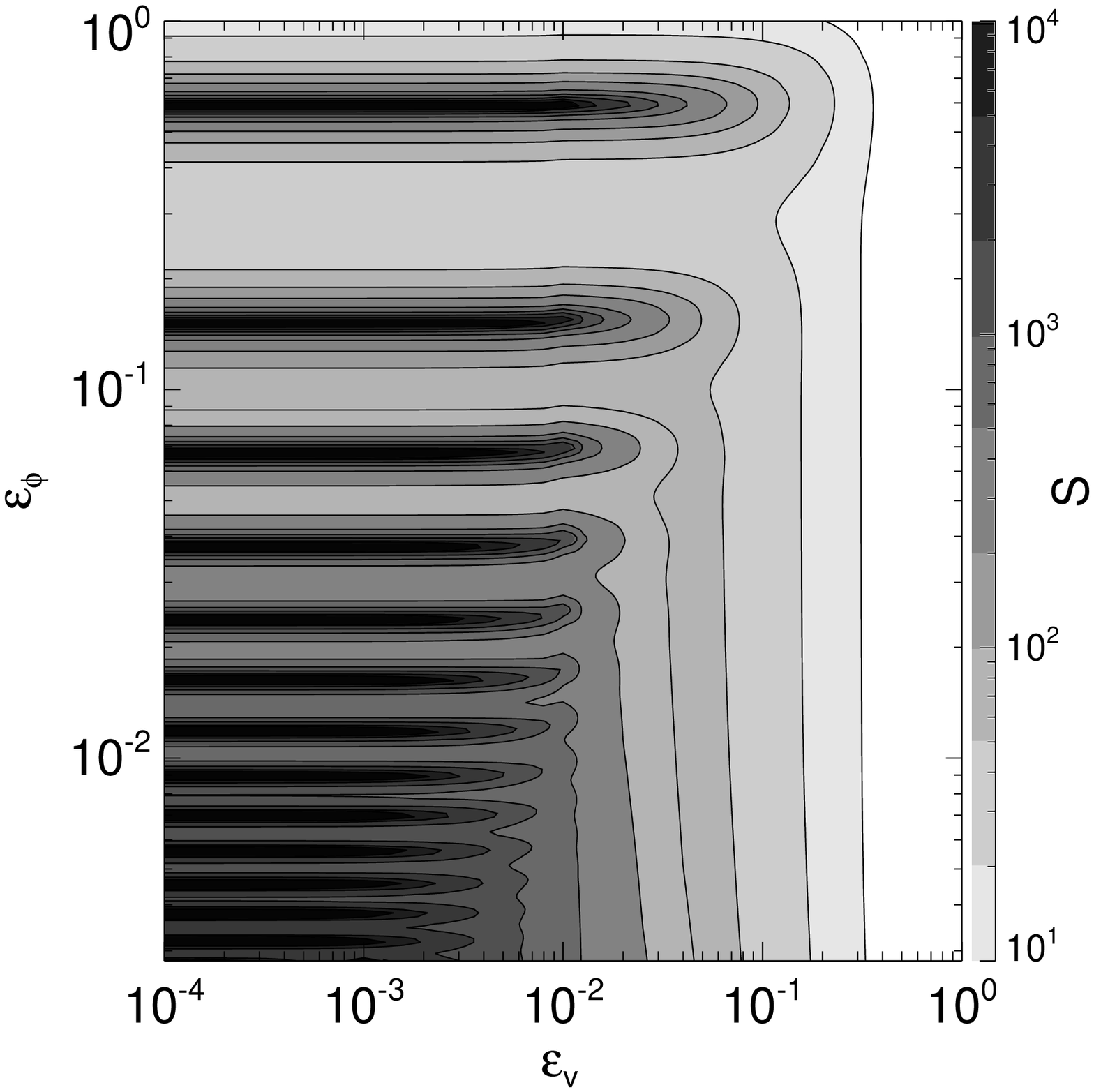}  
  \includegraphics[width=0.32\textwidth]{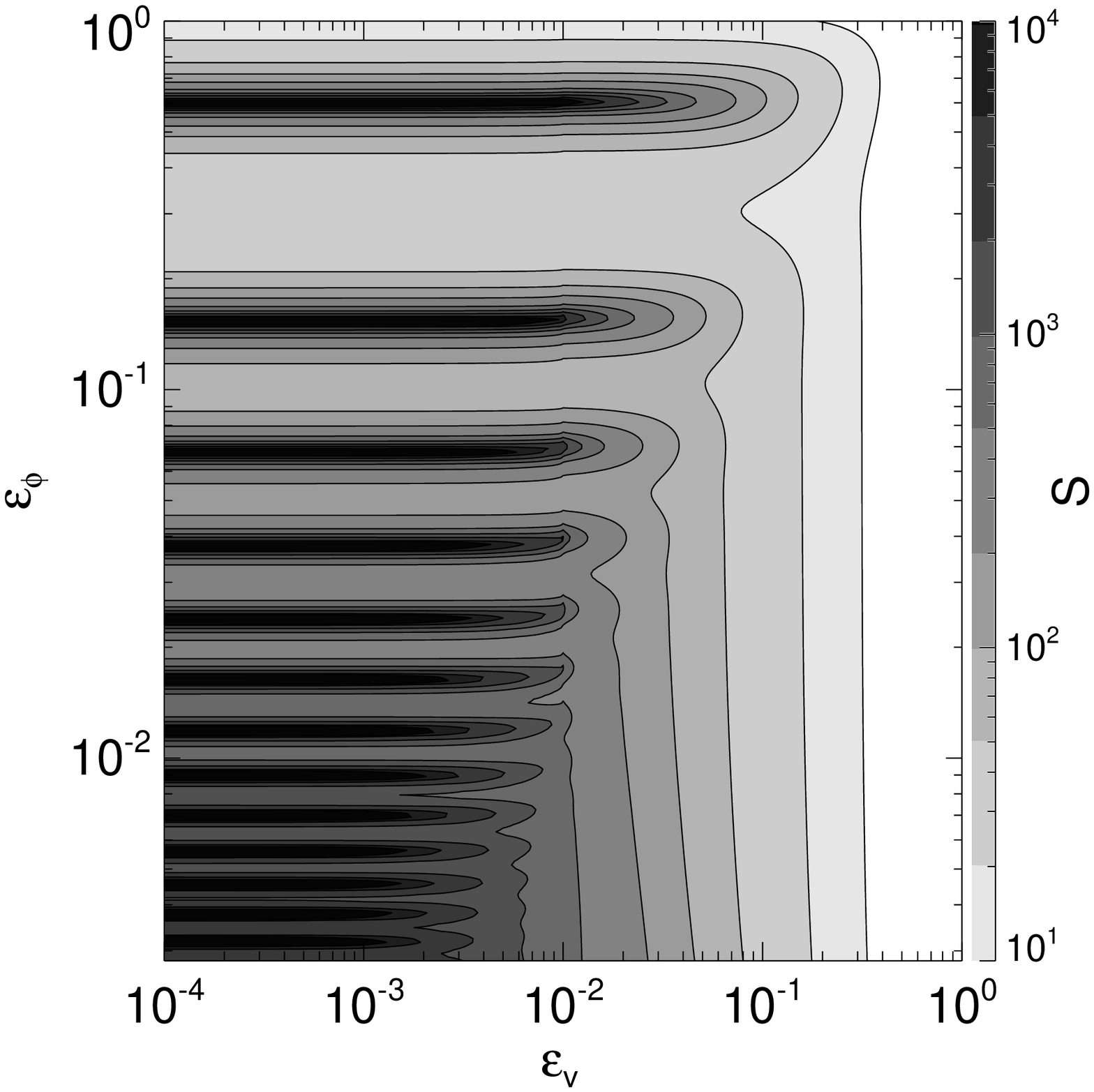}
  \includegraphics[width=0.32\textwidth]{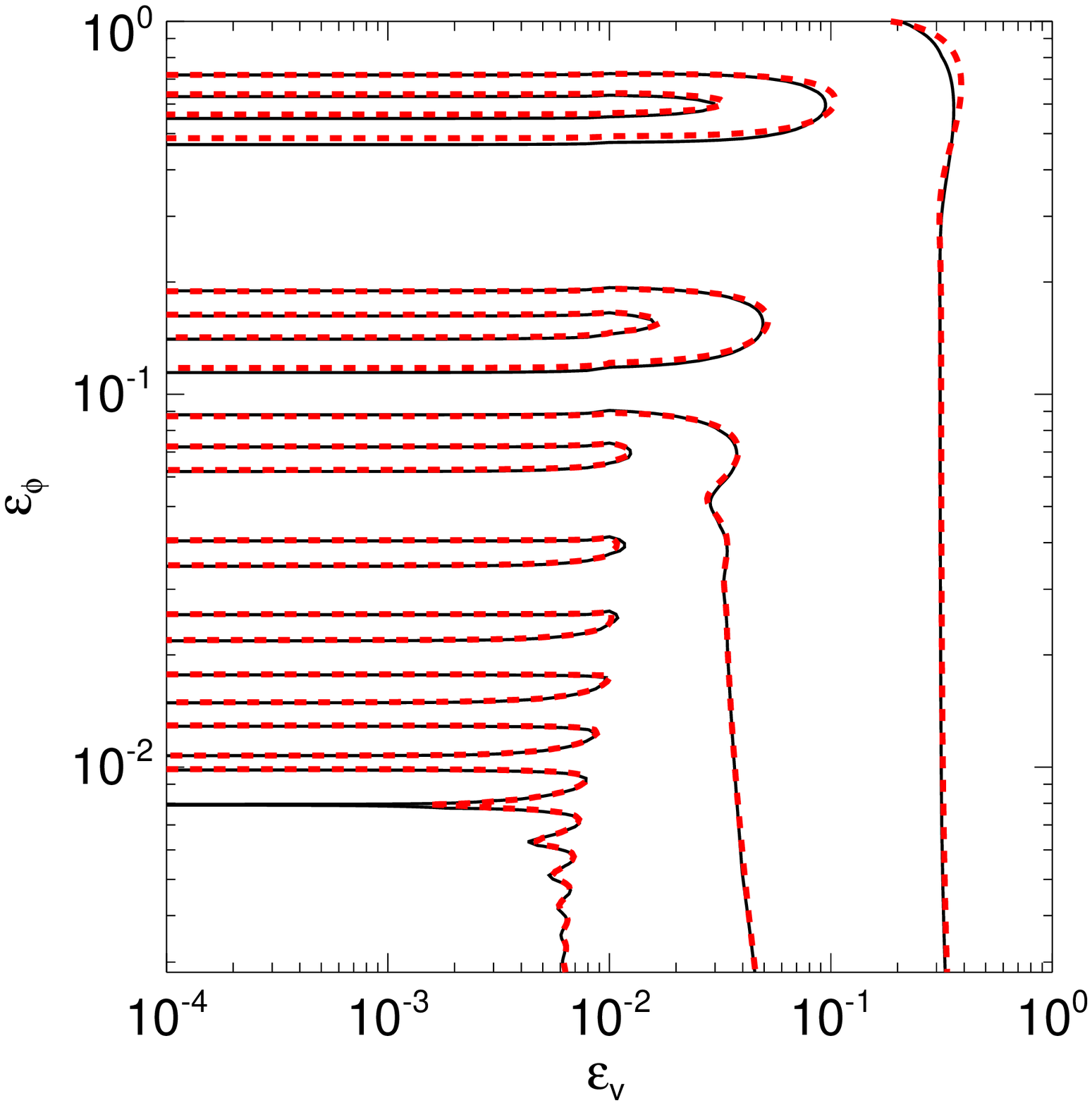} \\  
  \includegraphics[width=0.32\textwidth]{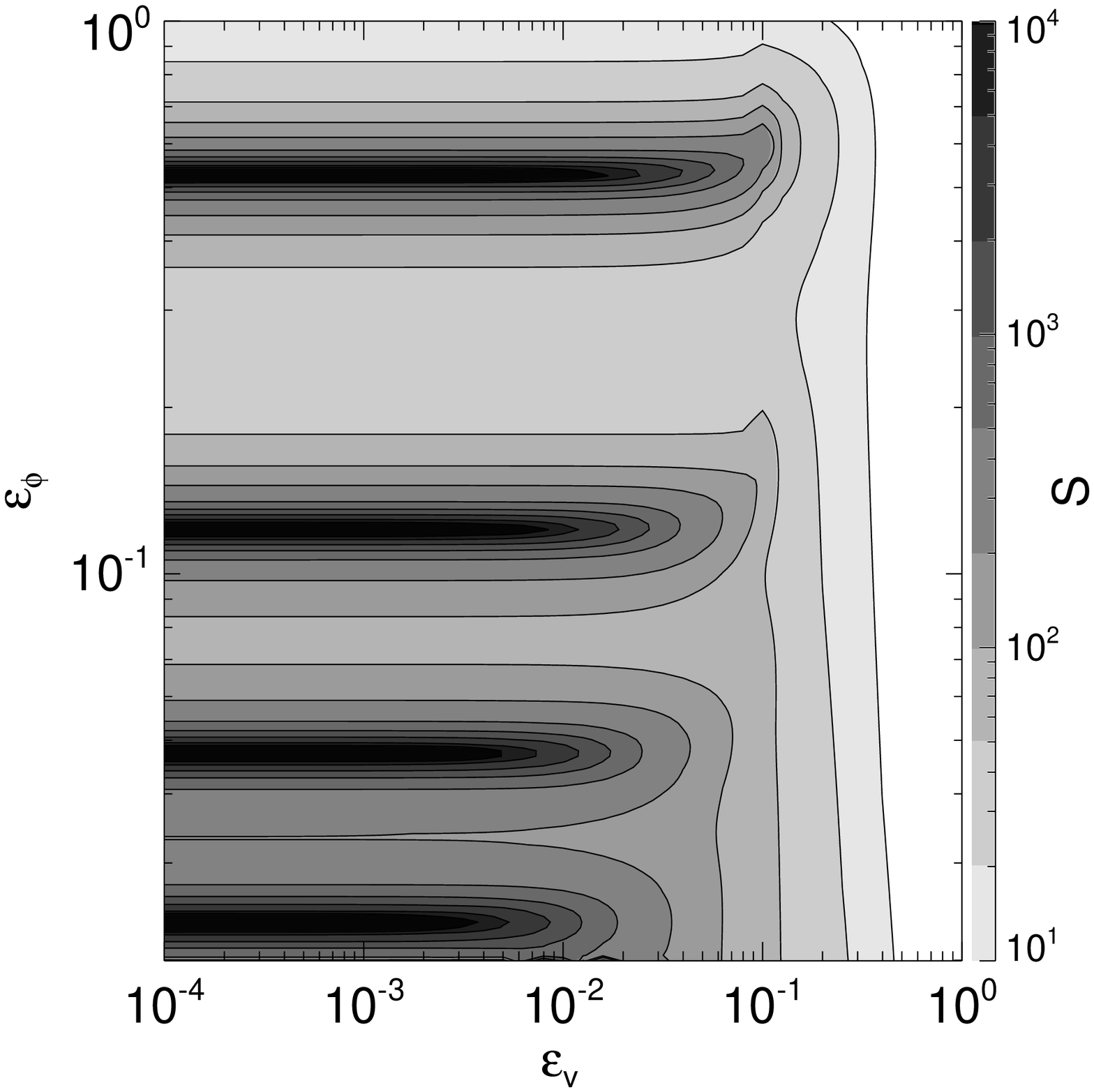}
  \includegraphics[width=0.32\textwidth]{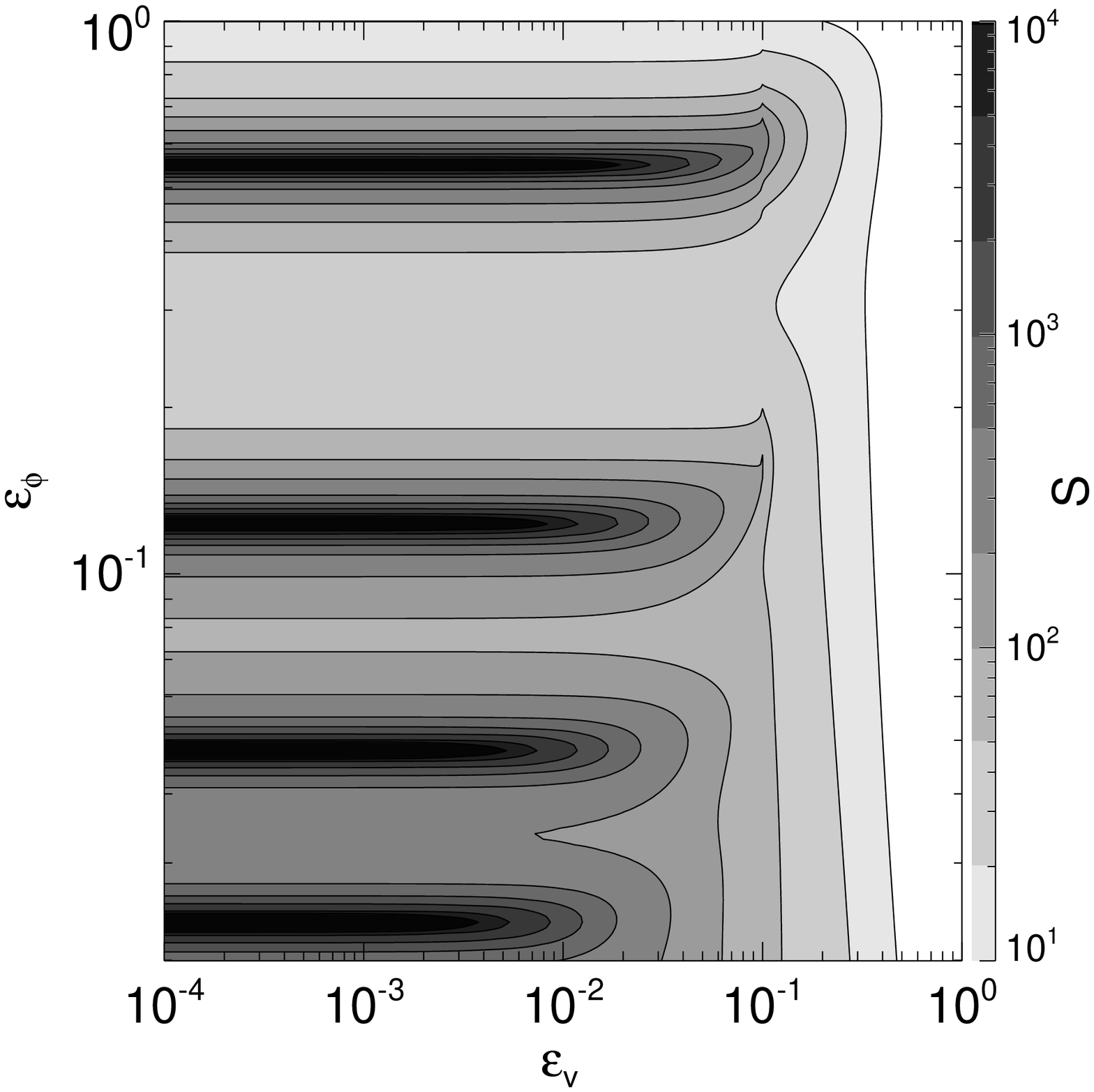}  
  \includegraphics[width=0.32\textwidth]{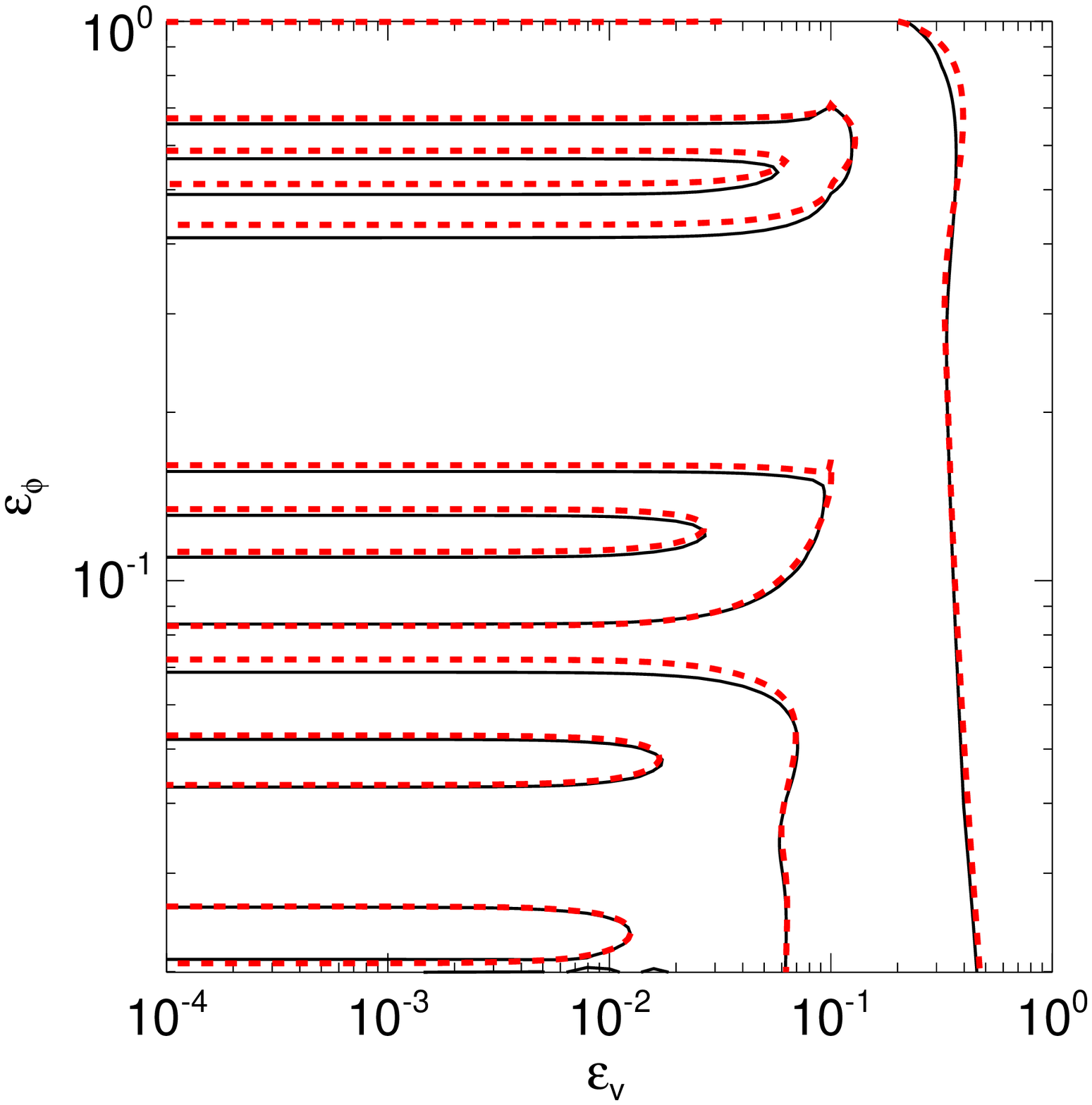}
\caption{\label{fig:numchecks} 
The Sommerfeld enhancement as a function of $\epsilon_\phi$ and $\epsilon_v$, computed both numerically and using the semi-analytic results derived in this work, for (\emph{top}) $\epsilon_\delta = 0$, (\emph{middle}) $\epsilon_\delta = 0.01$, and (\emph{bottom}) $\epsilon_\delta=0.1$. For purposes of comparison, we restrict the range of $\epsilon_\phi$ to regions where the numerical solutions to the Schr\"{o}dinger equation are stable. \emph{Left panel:} The enhancement computed by numerically solving the Schr\"{o}dinger equation using \texttt{Mathematica}, \emph{Middle panel:} The enhancement computed from Eq. \ref{eq:simple}, \emph{Right panel:} the $S=10, \, 100, \, 1000$ contours for the numeric calculation of $S$ (\emph{solid black line}) overlaid on the same contours for the approximate calculation (\emph{dashed red line}).
}}

\subsection{Magnitude of the non-resonant enhancement}
\label{sec:nonresonant}

For $\epsilon_v \gg \epsilon_\delta$, $\epsilon_v \gg \mu$, we obtain,
\[ S \approx \left( \frac{\pi}{\epsilon_v} \right) \frac{\sinh\left( \frac{2 \pi \epsilon_v}{\mu}\right)}{\cosh\left(\frac{2 \pi \epsilon_v}{\mu}\right) - \cos(2 \theta_-)} , \]
and furthermore the exponential terms dominate the oscillatory $\cos$ term, yielding $S \approx \pi / \epsilon_v$. Thus we recover the usual Coulomb enhancement in the limit where the mass splitting and mediator mass parameters are small compared to the velocity parameter $\epsilon_v$, as required.

If $\epsilon_v \gg \mu$ but $\epsilon_v < \epsilon_\delta$, on the other hand, then as previously the exponential terms dominate the oscillatory term, but now $S \approx 2 \pi / \epsilon_v$. So below the threshold for on-shell excitation of the upper state, the non-resonant enhancement is increased by a factor of 2. This effect is illustrated in Fig. \ref{fig:comparison}.

\FIGURE[h]{
  \includegraphics[width=0.45\textwidth]{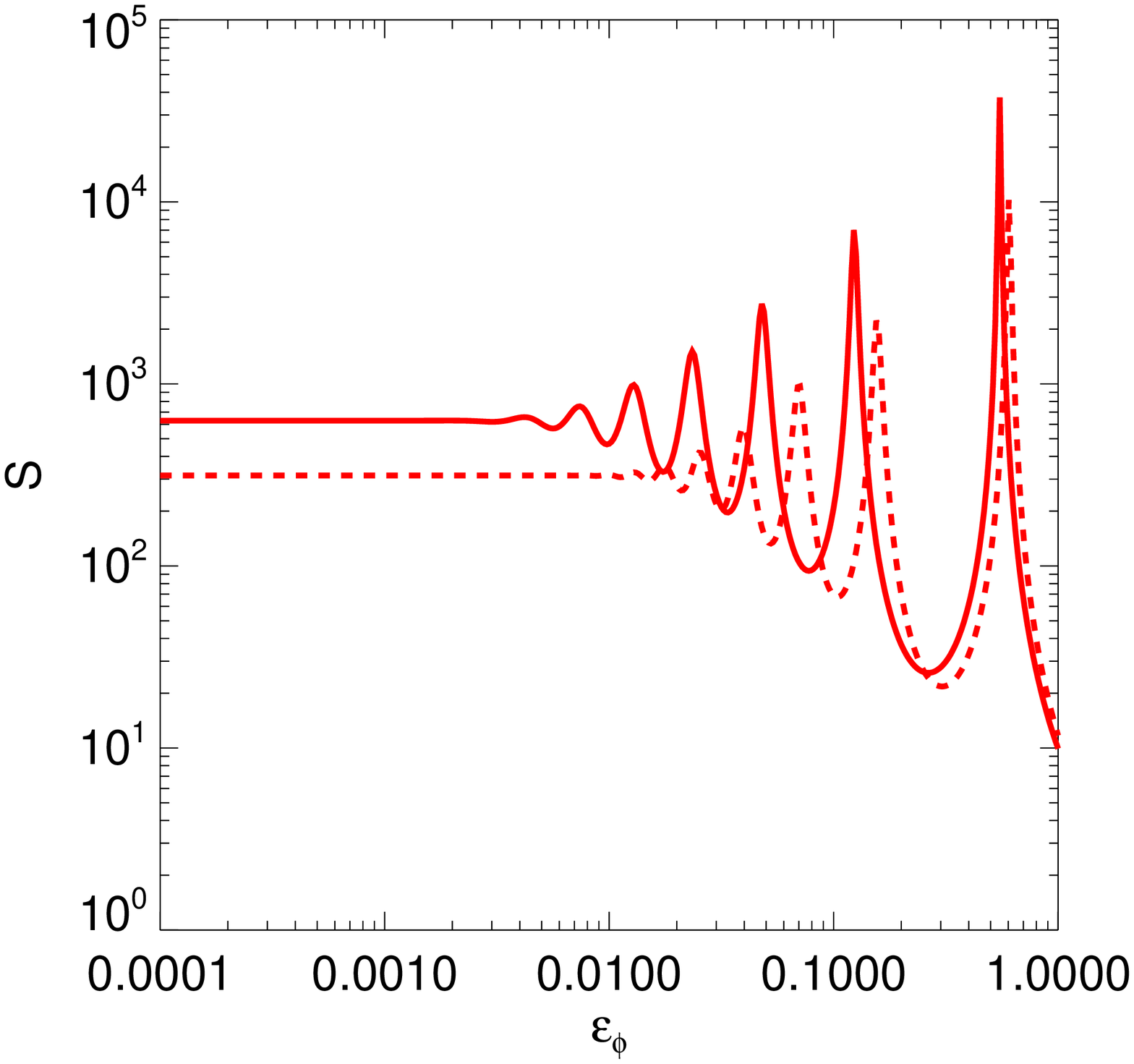}
  \includegraphics[width=0.45\textwidth]{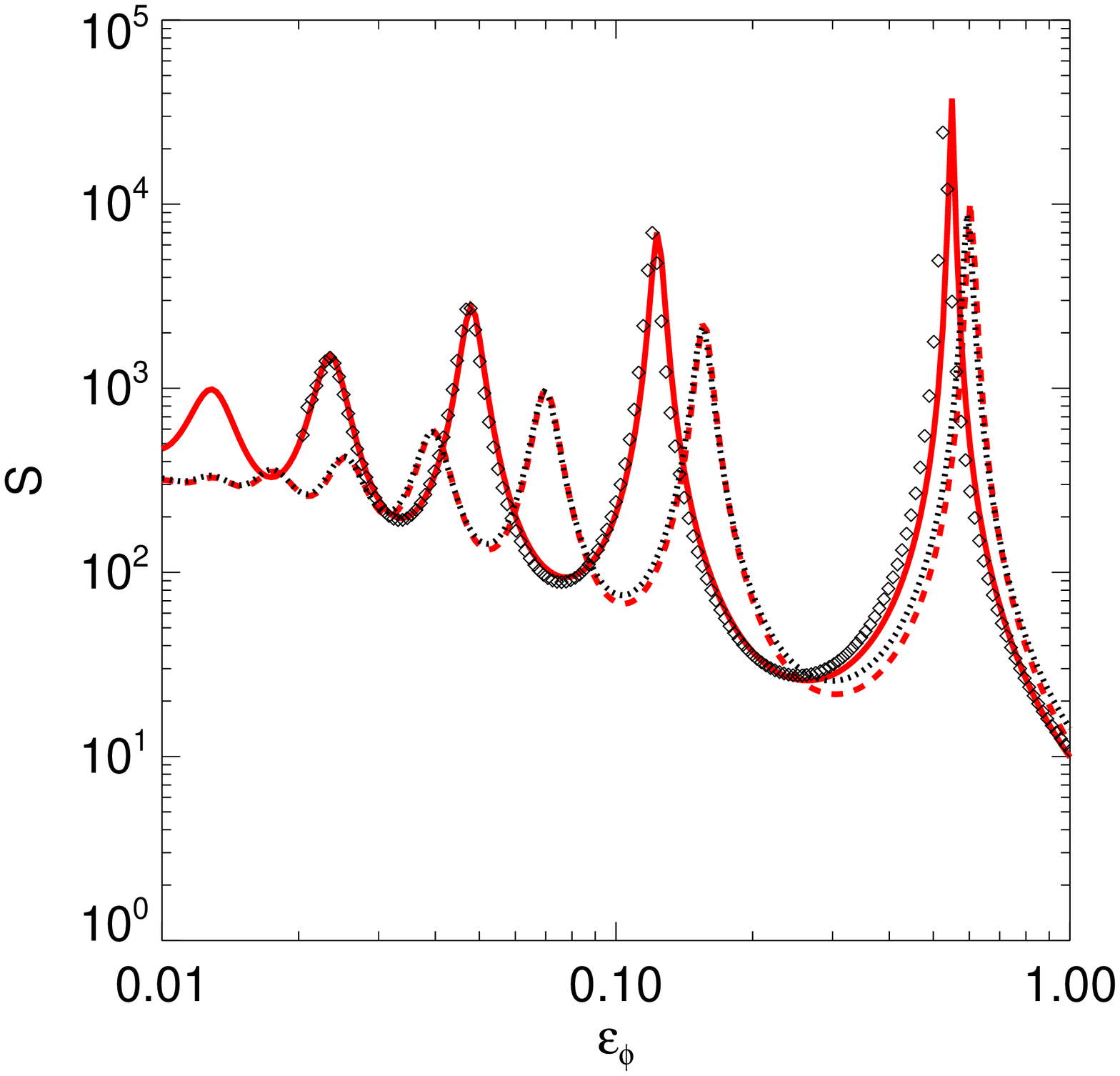} 
\caption{\label{fig:comparison} 
 The Sommerfeld enhancement as a function of $\epsilon_\phi$, for $\epsilon_v = 0.01$, and (\emph{solid line}) $\epsilon_\delta=0.1$, (\emph{dashed line}) $\epsilon_\delta=0$, using the approximate expressions described in the text. \emph{Left panel:} The left-hand side of the plot ($\epsilon_\phi \ll \epsilon_v < \epsilon_\delta $) demonstrates the increase in the non-resonant enhancement discussed in \S \ref{sec:nonresonant}, whereas the right-hand side ($\epsilon_v \lesssim \epsilon_\phi$) shows the higher, shifted resonances discussed in \S \ref{sec:resonances}. \emph{Right panel:} As the left panel, but also displaying the results of numerically solving the Schr\"{o}dinger equation, in the region of numerical stability. \emph{Black diamonds}: $\epsilon_\delta=0.1$, \emph{black dots:} $\epsilon_\delta=0$. The approximate solutions reproduce the numerical results well, and the predicted larger resonances and shifts in the resonance positions are clearly visible.}
}

When $\epsilon_v \lesssim \mu$, then up to a factor of the form $1/(1 - \cos \theta)$ describing the resonances (which will be discussed next), both above and below the excitation threshold, the enhancement saturates at the same value $S \sim 2 \pi^2 / \mu$. In the regime where Eq. \ref{eq:simple} is expected to be accurate ($r_T \sim 1/\epsilon_\phi$), $\mu \sim \epsilon_\phi$, and this is the usual saturation of the enhancement occurring when $\epsilon_v$ drops below $\epsilon_\phi$. As $\epsilon_\phi \rightarrow 0$, $\mu \rightarrow 0$ (this holds regardless of $\epsilon_\delta$, provided only that $\epsilon_\delta \lesssim 1$), and so the enhancement does not saturate (scaling as $1/\epsilon_v$ down to arbitrarily small velocities) \footnote{While our method may be expected to become inaccurate for $\epsilon_\phi \lesssim \epsilon_\delta^2/2$, this general statement -- that in the $\epsilon_\phi = 0$ limit, at low velocity $S$ scales as $1/\epsilon_v$, independent of $\epsilon_\delta$ provided $\epsilon_\delta \lesssim 1$ -- is expected to remain valid, based on matching the WKB solution in the small-$r$ region onto the exact solution for an effective $1/r^2$ potential in the large-$r$ region.}.

\subsection{Resonance behavior}
\label{sec:resonances}

When $\epsilon_v$ falls below $\epsilon_\delta$, the resonance structure described by Eq. \ref{eq:simple} changes in two major ways. First, the heights of the resonances change: where $\epsilon_v < \epsilon_\delta$, the resonances can be significantly larger than in the zero-$\delta$ case. This can be seen simply by examining the $\cosh$ term in the denominator of Eq. \ref{eq:simple}, where $\epsilon_v \ll \mu$: the height of each resonance at its peak is controlled by $1/(\cosh(\epsilon_v \pi / \mu) - 1) \approx (\mu/\pi \epsilon_v)^2$ in the below-threshold case, and $1/(\cosh((\epsilon_v + \sqrt{\epsilon_v^2 - \epsilon_\delta^2}) \pi / \mu) - 1) \approx (\mu/\pi(\epsilon_v + \sqrt{\epsilon_v^2 - \epsilon_\delta^2}))^2$ above threshold. Thus we expect the resonances to be $\sim 4 \times$ higher in the below-threshold case compared to the zero-$\delta$ limit, as illustrated in Fig. \ref{fig:comparison}.

The resonances also shift to different values of $\epsilon_\phi$ when $\epsilon_v$ falls below $\epsilon_\delta$. We can read off the resonance locations from the semi-analytic solution: above threshold they occur when $\theta_- = n \pi$, $n \in \mathcal{Z}$, whereas below threshold they shift to satisfy $\theta_- - \pi \sqrt{\epsilon_\delta^2 - \epsilon_v^2}/2 \mu = n \pi$, $n \in \mathcal{Z}$. To estimate the resonance locations we need an analytic estimate for the integral $\theta_-$. 

In the range $1 \gtrsim \epsilon_\phi \gtrsim \epsilon_\delta$, $\epsilon_v \lesssim \mu$,  $\theta_-$ is very well approximated\footnote{This expression was derived by noting that in this regime the $\int^{r_M}_0 \sqrt{\lambda_-} dr$ contribution to $\theta_-$ dominates, and setting $\lambda_- \approx -V(r)$; the estimate was then checked numerically against the full integral form of $\theta_-$.} by $-\sqrt{2 \pi/\epsilon_\phi}$. The condition that $\epsilon_v \lesssim \mu$ is required for significant resonances (since otherwise the oscillatory terms in Eq. \ref{eq:simple} are entirely dominated by the exponential terms), and since the height of the resonances is proportional to $\mu^2 \sim \epsilon_\phi^2$, the most pronounced resonances occur at large $\epsilon_\phi$ where this simple estimate for $\theta_-$ is valid (for $\epsilon_\phi \lesssim \epsilon_\delta$ there are significant corrections to this expression, but no especially simple form has been found).

Employing this estimate for $\theta_-$, we find that above threshold, resonances occur where $\epsilon_\phi \approx (2/\pi)/n^2$, $n \in \mathcal{N}$. Below threshold, we write $\mu = c \epsilon_\phi$ where $c$ is a slowly varying $\mathcal{O}(1)$ function of $\epsilon_\phi$, and approximate $c$ = constant. Solving the resulting quadratic equation for the resonances at $\epsilon_v = 0$, we obtain,
\[ \sqrt{\frac{2 \pi}{\epsilon_\phi}} - \pi \frac{\epsilon_\delta}{2 c \epsilon_\phi} = n \pi \Rightarrow \epsilon_\phi = \frac{1}{2 n^2} \left(\frac{2}{\pi} - \frac{n \epsilon_\delta}{c} \pm \frac{2}{\pi} \sqrt{1 - n\pi \epsilon_\delta/c} \right). \]

It appears that this equation imposes an upper bound on $n$: however, recall that we have assumed $\epsilon_\phi \gtrsim \epsilon_\delta$. If the term under the square root is to approach zero, this corresponds to $n \sim c / \pi \epsilon_\delta$, and $\epsilon_\phi \sim 1/(2 \pi n_\mathrm{max}^2) \sim (\pi/2) \epsilon_\delta^2/c^2 \lesssim \epsilon_\delta$. Thus by virtue of our previous assumptions, we cannot probe the large-$n$ limit: we will restrict ourselves to the case where $n \pi \epsilon_\delta / c \ll 1$. In this regime, we can approximate the solutions for the resonance positions as:
\[\epsilon_\phi = \frac{2}{\pi} \frac{1}{n^2} - \frac{\epsilon_\delta}{c n}, \, \frac{1}{\pi n^2} \mathcal{O}\left(\left(n \pi \epsilon_\delta / c\right)^2\right). \]
Only the first set of solutions lies in the $\epsilon_\phi \gtrsim \epsilon_\delta$ regime that we have assumed, so the effect is to shift the existing resonances downward in $\epsilon_\phi$ by $\epsilon_\delta/c n$ (note that relative to the resonance position, this corresponds to a fractional shift of $\pi n \epsilon_\delta/2 c$, i.e. the fractional shift grows with increasing $n$). Fig. \ref{fig:comparison} clearly shows the increasing fractional shift to lower values of $\epsilon_\phi$ with increasing $n$.

\subsection{Behavior at the $\epsilon_v = \epsilon_\delta$ threshold}

In the $\delta=0$ limit, $S$ is a monotonic function of $\epsilon_v$. This no longer holds true when $\delta \ne 0$: the position of the resonances varies as a function of $\epsilon_v$, and this effect can give rise to noticeable spikes in the enhancement around the threshold velocity $\epsilon_v = \epsilon_\delta$ (where the resonance positions are most dependent on $\epsilon_v$; for $\epsilon_v$ far from  $\epsilon_\delta$, the resonance positions are largely independent of $\epsilon_v$). In particular, the Sommerfeld enhancement near threshold may be even larger than the saturated enhancement at low velocity. In the context of DM annihilation, this effect could alleviate tension with early-universe constraints on the rate of dark matter annihilation (e.g. \cite{Belikov:2009qx,Profumo:2009uf,Galli:2009zc,Slatyer:2009yq,Huetsi:2009ex,Cirelli:2009bb}), since in the early universe the DM is extremely slow-moving, while in the present-day Galactic halo $\epsilon_v$ is of the same order as $\epsilon_\delta$ for phenomenologically interesting mass splittings \footnote{For example, for a $1$ TeV WIMP with $100$ keV mass splitting, $\epsilon_v \sim \epsilon_\delta$ corresponds to $v/c \sim 130$ km/s, which is very close to estimates for the one-particle rms velocity of the WIMPs \cite{Governato:2007}.}. To investigate this possibility we examine the ratio,
\[R_\mathrm{thres} = \frac{S(\epsilon_v=\epsilon_\delta)}{S(\epsilon_v = 0)} = \frac{\mu}{\pi \epsilon_\delta} \frac{\sinh\left(\frac{\pi \epsilon_\delta}{\mu}\right) \left(1 - \cos \left(\frac{\pi \epsilon_\delta}{\mu} + 2 \theta_-(\epsilon_v=0) \right) \right)}{\cosh\left(\frac{\pi \epsilon_\delta}{\mu} \right) - \cos \left(2 \theta_-(\epsilon_v = \epsilon_\delta) \right)}. \]

If $\pi \epsilon_\delta \gtrsim \mu$, then the exponential terms dominate, and $R_\mathrm{thres} \approx (\mu/\pi \epsilon_\delta) \left(1 - \cos \left(\frac{\pi \epsilon_\delta}{\mu} + 2 \theta_-(\epsilon_v=0) \right) \right)$. We see that $R_\mathrm{thres} \le 2 \mu / \pi \epsilon_\delta \lesssim 2$, and the maximal value of $R_\mathrm{thres}$ will occur when $\mu \sim \pi \epsilon_\delta$ and $\pi \epsilon_\delta / \mu + 2 \theta_-(\epsilon_v = 0)$ is as close as possible to $(2 n - 1) \pi$, $n \in \mathcal{N}$. An example where $R_\mathrm{thres}$ attains a large value is shown in Fig. \ref{fig:thresres}. Of course, in realistic applications it will be necessary to integrate the enhancement over some velocity distribution, which will tend to smooth out the resonance: the effect of smoothing by a Maxwell-Boltzmann distribution is shown in the right-hand panel of Fig. \ref{fig:thresres}.

\FIGURE[h]{
\includegraphics[width=0.3\textwidth]{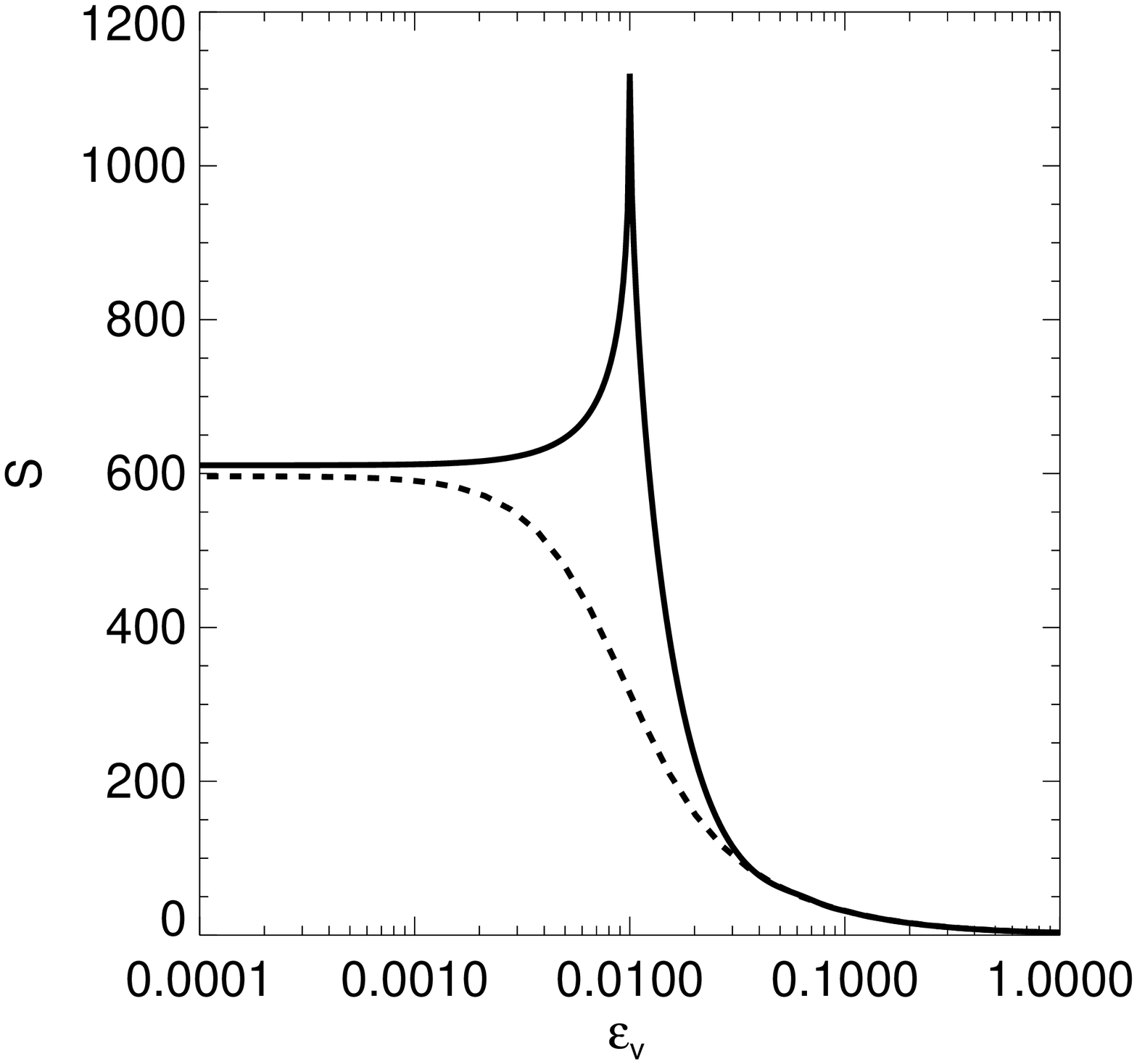}  
\includegraphics[width=0.3\textwidth]{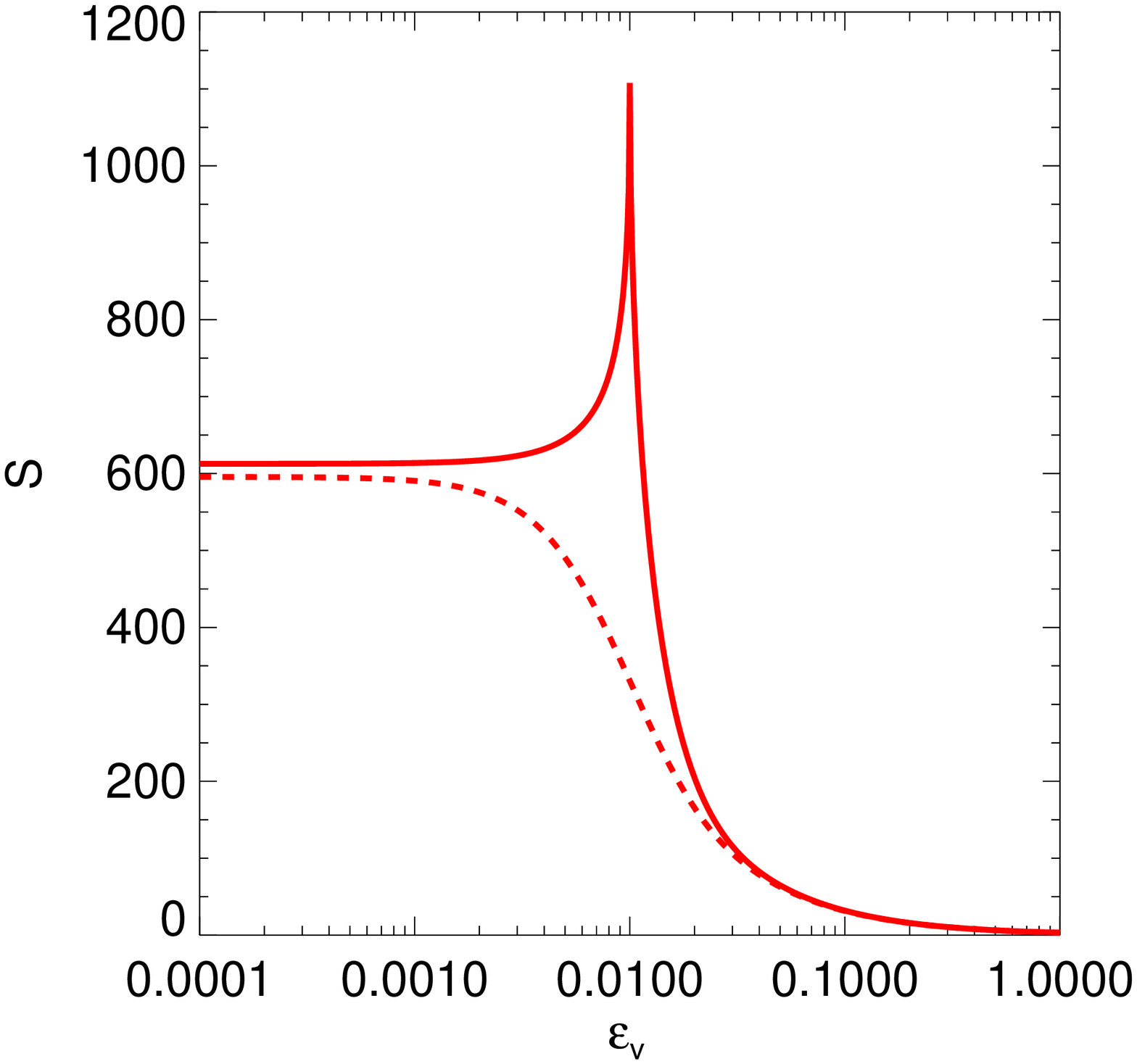}
\includegraphics[width=0.3\textwidth]{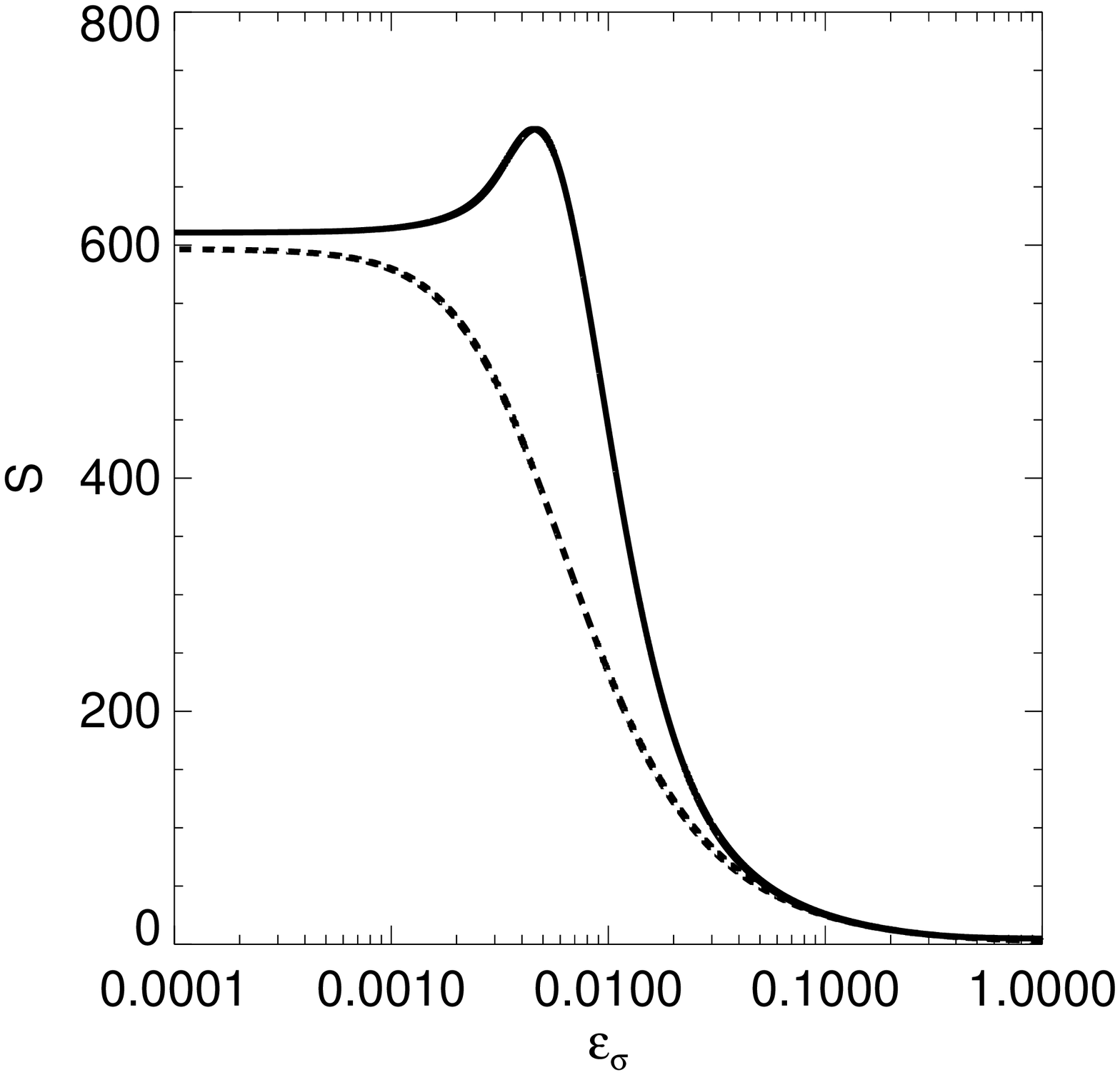}
\caption{\label{fig:thresres} 
Examples of non-monotonicity of the Sommerfeld enhancement as a function of $\epsilon_v$. \emph{Left panel:} the enhancement computed by numerically solving the Schr\"{o}dinger equation, for (\emph{solid line}) $\epsilon_\delta=0.01$, $\epsilon_\phi = 0.04$, (\emph{dashed line}) $\epsilon_\delta=0$, $\epsilon_\phi = 0.043$.  The slightly different values of $\epsilon_\phi$ are chosen to give the same behavior in the $\epsilon_v \rightarrow 0$ limit, to show the qualitatively different behavior at intermediate velocities for models which are similar at very low and very high velocities.  \emph{Middle panel:} the enhancement computed using the approximate results derived in this work, for  (\emph{solid line}) $\epsilon_\delta=0.01$, $\epsilon_\phi = 0.0414$, (\emph{dashed line}) $\epsilon_\delta=0$, $\epsilon_\phi = 0.0433$. The marginally different parameter choices for the analytic approximation and the numerical calculation reflect the small degree to which the approximation shifts the resonances; in both cases the parameters were chosen to demonstrate a clear enhancement at $\epsilon_v \sim \epsilon_\delta$. \emph{Right panel:} the effect of smoothing the velocity-dependent resonance (using the numerical results, and the same parameters as the left-hand panel) with a Maxwell-Boltzmann distribution $\epsilon_v^2 e^{-\epsilon_v^2/2 \epsilon_\sigma^2}$. The enhancement is plotted as a function of the dimensionless velocity dispersion $\epsilon_\sigma$.}
}

In the opposite limit, where $\pi \epsilon_\delta \ll \mu$,
\[ R_\mathrm{thres} \approx \frac{1 - \cos\left(\frac{\pi \epsilon_\delta}{\mu} + 2 \theta_-(\epsilon_v=0)\right)}{\left(1 + \left(\frac{\pi \epsilon_\delta}{\mu}\right)^2 - \cos\left(2 \theta_-(\epsilon_v=\epsilon_\delta) \right) \right)}. \]
There is the potential for a large enhancement if $\cos\left(2 \theta_-(\epsilon_v=\epsilon_\delta) \right) \approx 1$, leaving only a small term in the denominator. However, at lowest order $\theta_-$ is independent of $\epsilon_v$, so the numerator would then be zero at lowest order. Consequently, in order to determine $R_\mathrm{thres}$ in this limit we must compute the next-order term in $\theta_-$ for $\epsilon_v = 0$ and $\epsilon_v = \epsilon_\delta$.

Away from the transition region (where $V$ and $\tilde{V}$ cancel out in $\theta_-$), we can approximate $\sqrt{\lambda_-} = i \sqrt{V} \pm i \epsilon_\delta^2 / 4 \sqrt{V}$, where the $+$ sign corresponds to $\epsilon_v = \epsilon_\delta$ and the $-$ sign to $\epsilon_v = 0$. An identical expression holds for $\sqrt{\tilde{\lambda_-}}$ with the replacement $V \rightarrow \tilde{V}$. Thus,
\[ \theta_-(\epsilon_v=\epsilon_\delta) - \theta_-(\epsilon_v=0) = \frac{\epsilon_\delta^2}{2} \left( \int^{r_T}_{r_S} 1 / \sqrt{\tilde{V}} dr - \int^{r_T}_0 1/\sqrt{V} dr \right).\]
The first integral evaluates to $2 \sqrt{2}/(\mu \epsilon_\delta)$ (since $\tilde{V}(r_T) = \epsilon_\delta^2/2$), whereas the second scales like $\epsilon_\phi^{-3/2}$; assuming $\mu \sim \epsilon_\phi$, the first term dominates provided $\epsilon_\delta \lesssim \sqrt{\epsilon_\phi}$. Thus we have,
\[ \theta_-(\epsilon_v=\epsilon_\delta) \approx \theta_-(\epsilon_v=0) + \sqrt{2} \epsilon_\delta / \mu, \]
and consequently,
\[ R_\mathrm{thres} \approx \frac{1 - \cos\left(\frac{(\pi - 2 \sqrt{2}) \epsilon_\delta}{\mu} + 2 \theta_-(\epsilon_v=\epsilon_\delta)\right)}{\left(1 + \left(\frac{\pi \epsilon_\delta}{\mu}\right)^2 - \cos\left(2 \theta_-(\epsilon_v=\epsilon_\delta) \right) \right)}. \]
We see that as $\epsilon_\delta \rightarrow 0$, $R_\mathrm{thres} \rightarrow 1$, and if $\cos(2 \theta_-(\epsilon_v=\epsilon_\delta)) \rightarrow 1$, $R_\mathrm{thres} \rightarrow (1 - 2 \sqrt{2}/\pi)^2 < 1$. As well as these limits, it can be numerically checked that everywhere in this region of parameter space $R_\mathrm{thres} \le 1$. Consequently, $R_\mathrm{thres}$ is robustly bounded above by $R_\mathrm{thres} \le 2$.

\subsection{The $\epsilon_\delta \rightarrow 0$ limit and comparison to previous results}

In the case where the mass splitting is zero, $\epsilon_\delta=0$, Eq. \ref{eq:simple} simplifies to,
\[ S = \left( \frac{2 \pi}{\epsilon_v} \right) \frac{\sinh\left(\frac{\pi \epsilon_v}{\mu}\right)\cosh\left(\frac{\pi \epsilon_v}{\mu}\right)}{\cosh\left(\frac{2 \pi \epsilon_v}{\mu}\right) - \cos(2 \theta_-)} = \left( \frac{\pi}{\epsilon_v} \right) \frac{\sinh\left( \frac{2 \pi \epsilon_v}{\mu}\right)}{\cosh\left(\frac{2 \pi \epsilon_v}{\mu}\right) - \cos(2 \theta_-)}. \]
The phase $\theta_-$ is approximately given by $\theta_- \approx -\sqrt{2 \pi / \epsilon_\phi}$ in the $\epsilon_v \rightarrow 0$ limit.

An approximate solution for the Sommerfeld enhancement due to a Yukawa potential was derived in \cite{Cassel:2009wt}, using the exact solution to the Schr\"{o}dinger equation with the Hulth\'{e}n potential \cite{1402-4896-76-5-003, MarchRussell:2008tu}. In terms of the dimensionless parameters used in this work, the expression for the $s$-wave enhancement derived in \cite{Cassel:2009wt} can be written as,
\begin{equation} S = \left(\frac{\pi}{\epsilon_v} \right) \frac{\sinh\left(\frac{2 \pi \epsilon_v}{\epsilon_\phi^*} \right)}{\cosh\left(\frac{2 \pi \epsilon_v}{\epsilon_\phi^*} \right) - \cos\left(2 \pi \sqrt{\frac{1}{\epsilon_\phi^*} - \frac{\epsilon_v^2}{\epsilon_\phi^{* 2}}} \right)}, \label{eq:cassel} \end{equation}
where $\epsilon_\phi^*$ is a parameter of the exactly solvable potential used to approximate the Yukawa potential, and set to be $\epsilon_\phi^* = (\pi^2/6) \epsilon_\phi$ in \cite{Cassel:2009wt}. We see that the two expressions are identical, with the replacements $\mu \leftrightarrow \epsilon_\phi^*$ and $\theta_- \leftrightarrow \pi \sqrt{\frac{1}{\epsilon_\phi^*} - \frac{\epsilon_v^2}{\epsilon_\phi^{* 2}}}$. At low velocities the resonance positions are set by the argument of the $\cos$ term, which is $\sim 2 \pi / \sqrt{\pi^2 \epsilon_\phi /6} = 2 \sqrt{6} / \sqrt{\epsilon_\phi}$ in Eq. \ref{eq:cassel} and $\sim 2 \sqrt{2 \pi} / \sqrt{\epsilon_\phi}$ in our approximate solution, in close agreement. Taking $r_M \sim \epsilon_\phi$ in Eq. \ref{eq:mu} yields $\mu \sim (1/2) (1 + \sqrt{5}) \epsilon_\phi$, which is close to the $\pi^2 \epsilon_\phi/6$ value employed in \cite{Cassel:2009wt} (however, $\mu/\epsilon_\phi$ is only approximately constant, and as $\epsilon_\phi$ varies some disagreement between our formula and that of \cite{Cassel:2009wt} will ensue). The solution of \cite{Cassel:2009wt} did not make the approximation that $\epsilon_\phi, \epsilon_v \ll 1$, and thus is superior in the region where these assumptions break down and the enhancement is small, but the two approximations are very similar (and both reproduce the numeric results accurately) whenever the enhancement is large.

\section{Conclusion}

We have developed a detailed approximate expression for the $s$-wave Sommerfeld enhancement in the presence of a mass splitting, in the case where only inelastic scattering between the interacting particles is possible at tree level. The $\delta \rightarrow 0$ limit of our results yields an accurate approximation for the enhancement due to an attractive scalar Yukawa potential. The enhancement cuts off at low velocity if the mass splitting $\delta \gtrsim \alpha^2 m_\chi$, the effective Bohr energy of the two-body state, but even for smaller mass splittings, interesting modifications of the enhancement relative to the $\delta \rightarrow 0$ case can occur. Additionally, we have derived simple expressions to facilitate computation of the Sommerfeld enhancement due to an arbitrary $N \times N$ matrix potential, where the $l$th partial wave dominates. 

In the special case where all elements of the annihilation matrix are identical, our expression for the $s$-wave enhancement takes a simple form,
\[ S = \frac{2 \pi }{\epsilon_v}\sinh\left(\frac{\epsilon_v \pi }{\mu}\right) \left\{ \begin{array}{cc} \frac{1}{\cosh\left(\epsilon_v \pi/\mu\right)-\cos\left(\sqrt{\epsilon_\delta^2-\epsilon_v^2} \pi /\mu+2 \theta_-\right)} & \quad \epsilon_v < \epsilon_\delta, \\ \\ \frac{\cosh\left(\left(\epsilon_v+\sqrt{-\epsilon_\delta^2+\epsilon_v^2}\right) \pi/2 \mu\right) \text{sech}\left(\left(\epsilon_v-\sqrt{-\epsilon_\delta^2+\epsilon_v^2}\right) \pi/2 \mu \right)}{ \cosh\left(\left(\epsilon_v+\sqrt{-\epsilon_\delta^2+\epsilon_v^2}\right) \pi /\mu\right)-\cos(2 \theta_-)} &\quad \epsilon_v > \epsilon_\delta. \end{array} \right. \]
where $\mu$ and $\theta_-$ are real functions which we have defined. This expression is valid provided $m_\phi \lesssim \alpha m_\chi$, $v/c \lesssim \alpha$, and $\delta \lesssim \alpha^2 m_\chi$, although it may become less accurate for $\delta \gtrsim \max(\alpha m_\phi, \, m_\chi (v/c)^2)$. If any of the three former conditions are violated, there is no significant enhancement ($S \sim \mathcal{O}(1)$). In the regions where this expression is expected to be valid, comparison to numerical results demonstrates that it accurately reproduces all significant features of the enhancement. Furthermore, while derived for the case where all elements of the annihilation matrix are identical, a simple rescaling of this result by the average of the elements of the annihilation matrix (normalized to the $\Gamma_{11}$ element) can accurately describe the enhancement for a general annihilation matrix.

Employing this approximate form for the enhancement, we have shown that below the threshold for on-shell production of the excited state, in comparison to the case with no mass splitting:
\begin{itemize}
\item The non-resonant, unsaturated enhancement is higher by a factor of $\sim 2$.
\item The positions of the resonances shift to lower values of $\epsilon_\phi$ ($m_\phi / \alpha m_\chi$) and become more widely spaced. The heights of the resonances increase by a factor of $\sim 4$.
\item The enhancement $S$ is no longer a monotonic function of velocity: close to the threshold for on-shell excitation of the higher-mass state, the enhancement may be as much as a factor of $\sim 2$ greater than its saturated value (i.e. its value in the limit $v \rightarrow 0$).
\end{itemize}

In the context of DM annihilation, if the kinetic energy of WIMPs in the neighborhood of the Earth is close to the threshold for excitation of the higher-mass state (which is true by construction for models which realize the iDM mechanism), increased near-threshold enhancement may slightly alleviate constraints on DM annihilation from the early universe, where the DM is extremely slow-moving. The generically larger values of the unsaturated non-resonant enhancement, and the shift in position of the resonances, may help provide sufficiently large enhancements to generate recently observed cosmic-ray excesses, particularly for lower-mass force carriers. 
\vskip 18pt 
\noindent {\bf Acknowledgements}: I am grateful to Nima Arkani-Hamed for suggesting this analysis, to Douglas Finkbeiner for many helpful discussions, and to Neal Weiner for pointing out useful references and the possibility of near-threshold resonances. I thank the anonymous referee for their useful feedback and suggestions; Lisa Goodenough, Rob Morris, David Poland, David Rosengarten, David Shih, Natalia Toro and Kathryn Zurek for helpful comments and conversations; and the NYU Center for Cosmology and Particle Physics and the Michigan Center for Theoretical Physics for their hospitality during the completion of this work. This work was partially supported by a Sir Keith Murdoch Fellowship from the American Australian Association.

\appendix

\section{The exact solution for an exponential potential}
\label{sec:expexactsol}

We will employ an exactly solvable exponential potential to analyze the non-adiabatic transition at $V(r) \sim \epsilon_\delta^2/2$. If we take $\tilde{V}(r) = V_0 e^{-\mu r}$, then the Schr\"{o}dinger equation
\begin{equation} \left( \begin{array}{c} \psi_1''(r) \\ \psi_2''(r) \end{array} \right) = \left( \begin{array}{cc} - \epsilon_v^2 & - \tilde{V}(r) \\ - \tilde{V}(r)  & \epsilon_\delta^2 - \epsilon_v^2 \end{array} \right) \left( \begin{array}{c} \psi_1(r) \\ \psi_2(r) \end{array} \right) \nonumber \end{equation}
can be solved exactly by the substitution \cite{PhysRevA.49.265} $z = V_0^2 e^{-2 \mu r}/16 \mu^4$, since then the 4th-order ODE for $\psi_1$ takes the form,
\begin{equation} - \prod_{n=1}^4 \left(z \frac{d}{dz}- b_n \right) \psi_1 + z \psi_1 = 0, \end{equation}
\[ b_1 = i \epsilon_v / 2 \mu, \, b_2 = -i \epsilon_v / 2 \mu, \, b_3 = 1/2 + i \sqrt{\epsilon_v^2 - \epsilon_\delta^2}/2 \mu, \, b_4 = 1/2 - i \sqrt{\epsilon_v^2 - \epsilon_\delta^2}/2 \mu, \]
which is exactly solvable in terms of hypergeometric functions. $\psi_2$ can be deduced immediately from $\psi_1$, since $\psi_2 = (\psi_1'' + \epsilon_v^2 \psi_1)/\tilde{V}(r)$, and simplified using  the hypergeometric function identities,
\[(z\frac{d}{dz} + b_k - 1)\,_pF_q(a_1,\dots,a_p;b_1,\dots,b_k,\dots,b_q;z) = (b_k - 1)\,_pF_q(a_1,\dots,a_p;b_1,\dots,b_k-1,\dots,b_q;z)\]
\[\frac{d}{dz}\,_pF_q(a_1,\dots,a_p;b_1,\dots,b_q;z) = 
\frac{a_1\cdot\dots\cdot a_p}{b_1\cdot\dots\cdot b_q}\,_pF_q(a_1+1,\dots,a_p+1;b_1+1,\dots,b_q+1;z).\]

We find,
\begin{eqnarray}\psi_1 & = & C_2 z^{-\frac{i \epsilon_v}{2 \mu}}  \,_0F_3\left[\{\},\left\{1-\frac{i \epsilon_v}{\mu},\frac{1}{2}-\frac{i \epsilon_v}{2 \mu}-\frac{i \sqrt{-\epsilon_\delta^2+\epsilon_v^2}}{2 \mu},\frac{1}{2}-\frac{i \epsilon_v}{2 \mu}+\frac{i \sqrt{-\epsilon_\delta^2+\epsilon_v^2}}{2 \mu}\right\},z \right] \nonumber \\ 
& & + C_1 z^{\frac{i \epsilon_v}{2 \mu}}  \,_0F_3\left[\{\},\left\{1+\frac{i \epsilon_v}{\mu},\frac{1}{2}+\frac{i \epsilon_v}{2 \mu}-\frac{i \sqrt{-\epsilon_\delta^2+\epsilon_v^2}}{2 \mu},\frac{1}{2}+\frac{i \epsilon_v}{2 \mu}+\frac{i \sqrt{-\epsilon_\delta^2+\epsilon_v^2}}{2 \mu}\right\},z\right] \nonumber \\
& & + \sqrt{z} \left(C_4 z^{-\frac{i \sqrt{-\epsilon_\delta^2+\epsilon_v^2}}{2 \mu}}  \,_0F_3\left[\{\},\left\{\frac{3}{2}-\frac{i \epsilon_v}{2 \mu}-\frac{i \sqrt{-\epsilon_\delta^2+\epsilon_v^2}}{2 \mu},\frac{3}{2}+\frac{i \epsilon_v}{2 \mu}-\frac{i \sqrt{-\epsilon_\delta^2+\epsilon_v^2}}{2 \mu},1-\frac{i \sqrt{-\epsilon_\delta^2+\epsilon_v^2}}{\mu}\right\},z\right] \right. \nonumber \\ 
&&  \left. +  C_3 z^{\frac{i \sqrt{-\epsilon_\delta^2+\epsilon_v^2}}{2 \mu}} \,_0F_3\left[\{\},\left\{\frac{3}{2}-\frac{i \epsilon_v}{2 \mu}+\frac{i \sqrt{-\epsilon_\delta^2+\epsilon_v^2}}{2 \mu},\frac{3}{2}+\frac{i \epsilon_v}{2 \mu}+\frac{i \sqrt{-\epsilon_\delta^2+\epsilon_v^2}}{2 \mu},1+\frac{i \sqrt{-\epsilon_\delta^2+\epsilon_v^2}}{\mu}\right\},z\right]\right) \nonumber \end{eqnarray}
\begin{eqnarray}\psi_2 & = & - \sqrt{z} \left(  \frac{C_2 z^{-\frac{i \epsilon_v}{2 \mu}}}{\left(\frac{1}{2} - i \frac{\epsilon_v}{2 \mu}\right)^2 + \frac{\epsilon_v^2 - \epsilon_\delta^2}{4 \mu^2}} \,_0F_3\left[\{\},\left\{\frac{3}{2}-\frac{i \epsilon_v}{2 \mu}-\frac{i \sqrt{-\epsilon_\delta^2+\epsilon_v^2}}{2 \mu},\frac{3}{2}-\frac{i \epsilon_v}{2 \mu}+\frac{i \sqrt{-\epsilon_\delta^2+\epsilon_v^2}}{2 \mu},1-\frac{i \epsilon_v}{\mu}\right\},z\right]  \right. \nonumber \\ 
& & \left. +\frac{C_1 z^{\frac{i \epsilon_v}{2 \mu}}}{\left(\frac{1}{2} + i \frac{\epsilon_v}{2 \mu}\right)^2 + \frac{\epsilon_v^2 - \epsilon_\delta^2}{4 \mu^2}}  \,_0F_3\left[\{\},\left\{\frac{3}{2}+\frac{i \epsilon_v}{2 \mu}-\frac{i \sqrt{-\epsilon_\delta^2+\epsilon_v^2}}{2 \mu},\frac{3}{2}+\frac{i \epsilon_v}{2 \mu}+\frac{i \sqrt{-\epsilon_\delta^2+\epsilon_v^2}}{2 \mu},1+\frac{i \epsilon_v}{\mu}\right\},z\right] \right) \nonumber \\
& & - C_4 z^{-\frac{i \sqrt{-\epsilon_\delta^2+\epsilon_v^2}}{2 \mu}}\left(\frac{\epsilon_v^2}{4 \mu^2} + \left(\frac{1}{2} - i \frac{\sqrt{\epsilon_v^2 - \epsilon_\delta^2}}{2 \mu} \right)^2 \right) \nonumber \\ & & \times \,_0F_3\left[\{\},\left\{1-\frac{i \sqrt{\epsilon_v^2 - \epsilon_\delta^2}}{ \mu},\frac{1}{2}-\frac{i \epsilon_v}{2 \mu}-\frac{i \sqrt{-\epsilon_\delta^2+\epsilon_v^2}}{2 \mu},\frac{1}{2}+\frac{i \epsilon_v}{2 \mu}-\frac{i \sqrt{-\epsilon_\delta^2+\epsilon_v^2}}{2 \mu}\right\},z\right] \nonumber \\ 
&&  - C_3 z^{\frac{i \sqrt{-\epsilon_\delta^2+\epsilon_v^2}}{2 \mu}} \left(\frac{\epsilon_v^2}{4 \mu^2} + \left(\frac{1}{2} + i \frac{\sqrt{\epsilon_v^2 - \epsilon_\delta^2}}{2 \mu} \right)^2 \right) \nonumber \\ & & \times \,_0F_3\left[\{\},\left\{1+\frac{i \sqrt{\epsilon_v^2 - \epsilon_\delta^2}}{ \mu},\frac{1}{2}-\frac{i \epsilon_v}{2 \mu}+\frac{i \sqrt{-\epsilon_\delta^2+\epsilon_v^2}}{2 \mu},\frac{1}{2}+\frac{i \epsilon_v}{2 \mu}+\frac{i \sqrt{-\epsilon_\delta^2+\epsilon_v^2}}{2 \mu}\right\},z\right] \label{eq:exactsolapp} \end{eqnarray}

The hypergeometric functions asymptote to 1 as $r \rightarrow \infty$, so the prefactors determine their asymptotic behavior. We see that the four linearly independent solutions with coefficients $C_1, C_2, C_3$ and $C_4$ correspond to outgoing and ingoing spherical waves in the ground and excited states as $r \rightarrow \infty$:
\begin{itemize}
\item $C_1$ term: ingoing wave in the ground state, suppressed by $\sqrt{z}$ at large $r$ in the excited state,
\item $C_2$ term: outgoing wave in the ground state, suppressed by $\sqrt{z}$ at large $r$ in the excited state,
\item $C_3$ term: ingoing wave in the excited state, suppressed by $\sqrt{z}$ at large $r$ in the ground state,
\item $C_4$ term: outgoing wave in the excited state, suppressed by $\sqrt{z}$ at large $r$ in the ground state.
\end{itemize}

This solution can immediately be employed as an exactly solvable toy model for the Sommerfeld enhancement (similar to the finite square well, as mentioned in \cite{Hisano:2004ds}), but the resulting expression for the enhancement does not have a particularly enlightening form, so we do not discuss it further in this work. Instead, we employ this exact solution to describe the behavior of the wavefunction at large $r$.

\section{Matching of wavefunctions at $r \sim r_M$}
\label{sec:innermatch}

In the limit $z \rightarrow \infty$, the exact solution for the exponential potential (hereafter termed the ``exponential solution'' for brevity) takes a relatively simple form, determined by the asymptotics of the hypergeometric function:
\begin{equation} \,_0F_3\left[\{\},\left\{a,b,c\right\},z \right] \rightarrow \frac{\Gamma(a) \Gamma(b) \Gamma(c)}{2 (2\pi)^{3/2}} z^\gamma \left[ e^{4 z^{1/4}} + e^{2 \pi i \gamma} e^{4 i z^{1/4}} + e^{-2\pi i \gamma} e^{-4 i z^{1/4}} + \dots \right], \end{equation}
where $\gamma = (1/4) (3/2 - a - b - c)$ \cite{PhysRevA.62.062504}. There is also an exponentially suppressed term, of the form $\cos(4 \pi \gamma) e^{-4 z^{1/4}}$ \cite{0305-4470-31-2-034}; provided $\epsilon_v \lesssim \mu$, the imaginary part of $\gamma$ is small, and this term can be neglected (the effect of including this term has been tested, and it does not significantly modify our results). Where $\epsilon_v \gtrsim \mu$, the enhancement to annihilation takes a simple form, independent of the details of the matching procedure (i.e. terms involving the parameter $\theta_-$ are subdominant, and the dependence of $S$ on $\mu$ cancels out), and again this term can be neglected \footnote{Note that neglecting this term may give incorrect results if these wavefunctions are employed to study scattering in a similar system, in the $\epsilon_v \gtrsim \mu$ regime, since in the scattering problem the phase shifts induced by the potential are physically significant and depend on the matching procedure.}.

To determine the form of the exponential solution near the transition region, we use the WKB approximation to propagate the known asymptotic (large $z$) solution in to the transition region. We can then match onto the WKB solution for the inner region (\S \ref{sec:innerregionsol}).

For large $z$, the potential term dominates and the eigenstates are approximately $(-1,1)/\sqrt{2}$ (for the repulsed component) and $(1,1)/\sqrt{2}$ (for the attracted component). Writing the large-$z$ form of the exact solution (Eq. \ref{eq:exactsolapp}) in the form $\phi_+ \psi_+ + \phi_- \psi_-$, we can extract the solutions $\phi_\pm$ to the corresponding scalar Schr\"{o}dinger equations. 

Let us write the WKB solutions in the form,
\[\phi_\pm = \frac{1}{|\tilde{\lambda}_\pm|^{1/4}} \left( E_\pm e^{-\int^0_{r_S} \sqrt{\tilde{\lambda}_\pm(r')} dr'} e^{\int^r_{r_S} \sqrt{\tilde{\lambda}_\pm(r')} dr'} + F_\pm e^{\int^0_{r_S} \sqrt{\tilde{\lambda}_\pm(r')} dr'} e^{- \int^r_{r_S} \sqrt{\tilde{\lambda}_\pm(r')} dr'} \right), \]
where $\tilde{\lambda}_\pm = -\epsilon_v^2 + \epsilon_\delta^2/2 \pm \sqrt{(\epsilon_\delta^2/2)^2 + \tilde{V}(z)^2}$, and $r_S$ is chosen so that $V(r_S) \gg \epsilon_v^2, \, \epsilon_\delta^2$ (we choose this form to facilitate matching both onto the previously derived WKB solutions for the Yukawa potential, and onto the asymptotic forms of the exact solution for the exponential potential).

Choosing some $r > r_S$ that still lies within the potential-dominated region, the WKB solutions match smoothly onto the exact solution for the exponential potential, yielding the relations:
\begin{eqnarray} F_+ & = & -\frac{\sqrt{\mu}}{(2 \pi)^{3/2}} e^{-\int^0_{r_S} \sqrt{\tilde{\lambda}_+(r')} dr'} e^{4 z_S^{1/4}}  \\ & & \times \left(C_2 \Gamma\left[1-\frac{i \epsilon_v}{\mu}\right] \Gamma\left[\frac{-i \epsilon_v-i \sqrt{-\epsilon_\delta^2+\epsilon_v^2}}{2 \mu} + \frac{1}{2} \right] \Gamma\left[\frac{-i \epsilon_v+i \sqrt{-\epsilon_\delta^2+\epsilon_v^2}}{2 \mu} + \frac{1}{2}\right] \right. \nonumber \\
& & \left. + C_4 \Gamma\left[1-\frac{i \sqrt{-\epsilon_\delta^2+\epsilon_v^2}}{\mu}\right] \Gamma\left[-\frac{i \epsilon_v+i \sqrt{-\epsilon_\delta^2+\epsilon_v^2}}{2 \mu} + \frac{3}{2}\right] \Gamma\left[\frac{i \epsilon_v-i \sqrt{-\epsilon_\delta^2+\epsilon_v^2}}{2 \mu} + \frac{3}{2}\right] \right. \nonumber \\ 
& & \left. + C_1 \Gamma\left[1+\frac{i \epsilon_v}{\mu}\right] \Gamma\left[\frac{i \epsilon_v-i \sqrt{-\epsilon_\delta^2+\epsilon_v^2}}{2 \mu} + \frac{1}{2}\right] \Gamma\left[\frac{i \epsilon_v+i \sqrt{-\epsilon_\delta^2+\epsilon_v^2}}{2 \mu} + \frac{1}{2}\right] \right. \nonumber \\ 
& & \left. + i C_3  \Gamma\left[1+\frac{i \sqrt{-\epsilon_\delta^2+\epsilon_v^2}}{\mu}\right] \Gamma\left[\frac{-i \epsilon_v+i \sqrt{-\epsilon_\delta^2+\epsilon_v^2}}{2 \mu} + \frac{3}{2}\right] \Gamma\left[\frac{i \epsilon_v+i \sqrt{-\epsilon_\delta^2+\epsilon_v^2}}{2 \mu} + \frac{3}{2}\right] \right), \nonumber \label{eq:fplus} \end{eqnarray}
\begin{eqnarray} E_- & = & \frac{\sqrt{\mu}}{(2 \pi)^{3/2}} e^{\int^0_{r_S} \sqrt{\tilde{\lambda}_-(r')} dr'} e^{-4 i z_S^{1/4}}  \\ & & \times \left(C_2 e^{\frac{i \pi}{4} + \frac{\epsilon_v \pi }{\mu}} \Gamma\left[1-\frac{i \epsilon_v}{\mu}\right] \Gamma\left[\frac{-i \epsilon_v-i \sqrt{-\epsilon_\delta^2+\epsilon_v^2}}{2 \mu} + \frac{1}{2} \right] \Gamma\left[\frac{-i \epsilon_v+i \sqrt{-\epsilon_\delta^2+\epsilon_v^2}}{2 \mu} + \frac{1}{2}\right] \right. \nonumber \\
& & \left. + C_4 e^{\frac{5 \pi i}{4}+\frac{\pi \sqrt{-\epsilon_\delta^2+\epsilon_v^2}}{\mu}}\Gamma\left[1-\frac{i \sqrt{-\epsilon_\delta^2+\epsilon_v^2}}{\mu}\right] \Gamma\left[-\frac{i \epsilon_v+i \sqrt{-\epsilon_\delta^2+\epsilon_v^2}}{2 \mu} + \frac{3}{2}\right] \Gamma\left[\frac{i \epsilon_v-i \sqrt{-\epsilon_\delta^2+\epsilon_v^2}}{2 \mu} + \frac{3}{2}\right] \right. \nonumber \\
& & \left. + C_1 e^{\frac{i \pi}{4} - \frac{\epsilon_v \pi }{\mu}} \Gamma\left[1+\frac{i \epsilon_v}{\mu}\right] \Gamma\left[\frac{i \epsilon_v-i \sqrt{-\epsilon_\delta^2+\epsilon_v^2}}{2 \mu} + \frac{1}{2}\right] \Gamma\left[\frac{i \epsilon_v+i \sqrt{-\epsilon_\delta^2+\epsilon_v^2}}{2 \mu} + \frac{1}{2}\right] \right. \nonumber \\
& & \left. + i C_3 e^{\frac{5 \pi i}{4}- \frac{\pi \sqrt{-\epsilon_\delta^2+\epsilon_v^2}}{\mu}}  \Gamma\left[1+\frac{i \sqrt{-\epsilon_\delta^2+\epsilon_v^2}}{\mu}\right] \Gamma\left[\frac{-i \epsilon_v+i \sqrt{-\epsilon_\delta^2+\epsilon_v^2}}{2 \mu} + \frac{3}{2}\right] \Gamma\left[\frac{i \epsilon_v+i \sqrt{-\epsilon_\delta^2+\epsilon_v^2}}{2 \mu} + \frac{3}{2}\right] \right), \nonumber \label{eq:eminus} \end{eqnarray}

\begin{eqnarray} F_- & = & \frac{\sqrt{\mu}}{(2 \pi)^{3/2}} e^{-\int^0_{r_S} \sqrt{\tilde{\lambda}_+(r')} dr'} e^{4 i z_S^{1/4}} \\ & &\times  \left(C_2 e^{-\frac{i \pi}{4}-\frac{ \pi \epsilon_v}{\mu}} \Gamma\left[1-\frac{i \epsilon_v}{\mu}\right] \Gamma\left[\frac{-i \epsilon_v-i \sqrt{-\epsilon_\delta^2+\epsilon_v^2}}{2 \mu} + \frac{1}{2} \right] \Gamma\left[\frac{-i \epsilon_v+i \sqrt{-\epsilon_\delta^2+\epsilon_v^2}}{2 \mu} + \frac{1}{2}\right] \right. \nonumber \\
& & \left.  + C_4 e^{\frac{3 \pi i}{4} -\frac{\sqrt{-\epsilon_\delta^2+\epsilon_v^2} \pi }{\mu}}\Gamma\left[1-\frac{i \sqrt{-\epsilon_\delta^2+\epsilon_v^2}}{\mu}\right] \Gamma\left[-\frac{i \epsilon_v+i \sqrt{-\epsilon_\delta^2+\epsilon_v^2}}{2 \mu} + \frac{3}{2}\right] \Gamma\left[\frac{i \epsilon_v-i \sqrt{-\epsilon_\delta^2+\epsilon_v^2}}{2 \mu} + \frac{3}{2}\right] \right. \nonumber \\
& & \left. + C_1 e^{-\frac{i \pi}{4}+\frac{ \pi \epsilon_v}{\mu}} \Gamma\left[1+\frac{i \epsilon_v}{\mu}\right] \Gamma\left[\frac{i \epsilon_v-i \sqrt{-\epsilon_\delta^2+\epsilon_v^2}}{2 \mu} + \frac{1}{2}\right] \Gamma\left[\frac{i \epsilon_v+i \sqrt{-\epsilon_\delta^2+\epsilon_v^2}}{2 \mu} + \frac{1}{2} \right] \right. \nonumber \\
& & \left. + i C_3 e^{\frac{3 \pi i}{4} + \frac{\sqrt{-\epsilon_\delta^2+\epsilon_v^2} \pi }{\mu}} \Gamma\left[1+\frac{i \sqrt{-\epsilon_\delta^2+\epsilon_v^2}}{\mu}\right] \Gamma\left[\frac{-i \epsilon_v+i \sqrt{-\epsilon_\delta^2+\epsilon_v^2}}{2 \mu} + \frac{3}{2} \right] \Gamma\left[\frac{i \epsilon_v+i \sqrt{-\epsilon_\delta^2+\epsilon_v^2}}{2 \mu} + \frac{3}{2} \right] \right).\nonumber  \label{eq:fminus} \end{eqnarray}
As discussed above, we neglect the exponentially suppressed $E_+$ term.

Below threshold, there is no turning point in the WKB approximation for the repulsed eigenstate. At some matching point $r_M$, chosen so that the WKB approximation holds for both potentials and $V(r_M) \sim \tilde{V}(r_M)$, the matching equations become,
\begin{equation} E_\pm = \alpha_\pm e^{\int^{r_M}_0 \left( \sqrt{\lambda_\pm(r')} - \sqrt{\tilde{\lambda}_\pm(r')} \right) dr'}, \quad F_\pm  = \beta_\pm e^{-\int^{r_M}_0 \left( \sqrt{\lambda_\pm(r')} - \sqrt{\tilde{\lambda}_\pm(r')} \right) dr'}, \label{eq:matching}\end{equation}
\[ \alpha_- = - e^{i \pi/4} \frac{1}{2 \sqrt{\pi}} (A_- + i \pi \phi_-(0)), \quad \beta_- = e^{i \pi/4} \frac{i}{2 \sqrt{\pi}} \left(A_- - i \pi \phi_-(0) \right), \]
\[ \quad \alpha_+ = \frac{A_+}{2 \sqrt{\pi}}, \quad \beta_+ = \sqrt{\pi} \phi_+(0) - \frac{i}{2 \sqrt{\pi}} A_+. \]

Above threshold, suppose there is a turning point in the inner region, at $r = r^\dagger$, for the repulsive eigenstate of the exponential potential. Then the matching equations for the repulsive eigenstate are modified to give,
\begin{eqnarray} F_+ & = & \frac{i A_+}{2 \sqrt{\pi}} \left[ e^{2 \int^{r*}_0 \sqrt{\lambda_+} dr + \int^{r_M}_0 \left(\sqrt{\tilde{\lambda_+}} - \sqrt{\lambda_+}\right) dr} - e^{2 \int^{r^\dagger}_0 \sqrt{\tilde{\lambda_+}} dr + \int^{r_M}_0 \left(\sqrt{\lambda_+}-\sqrt{\tilde{\lambda_+}}  \right) dr} \right]  \\ & &
+ \frac{1}{2} \left( \sqrt{\pi} \phi_+(0) - \frac{i}{2 \sqrt{\pi}} A_+ \right) \left[ e^{\int^{r_M}_0 \left(\sqrt{\tilde{\lambda_+}} - \sqrt{\lambda_+}\right) dr} + e^{2 \int^{r^\dagger}_0 \sqrt{\tilde{\lambda_+}} dr -  2 \int^{r*}_0 \sqrt{\lambda_+} dr + \int^{r_M}_0 \left(\sqrt{\lambda_+}-\sqrt{\tilde{\lambda_+}}  \right) dr} \right] \nonumber \label{eq:fplusmatchingtp} \end{eqnarray}
\begin{eqnarray} E_+ & = & \frac{1}{2 i} \left\{ \frac{i A_+}{2 \sqrt{\pi}} \left[ e^{2 \int^{r*}_0 \sqrt{\lambda_+} dr - 2 \int^{r^\dagger}_0 \sqrt{\tilde{\lambda_+}} dr + \int^{r_M}_0 \left(\sqrt{\tilde{\lambda_+}} - \sqrt{\lambda_+}\right) dr} + e^{\int^{r_M}_0 \left(\sqrt{\lambda_+}-\sqrt{\tilde{\lambda_+}}  \right) dr} \right] \right.  \\ & &
\left. + \frac{1}{2} \left( \sqrt{\pi} \phi_+(0) - \frac{i}{2 \sqrt{\pi}} A_+ \right) \left[ e^{\int^{r_M}_0 \left(\sqrt{\tilde{\lambda_+}} - \sqrt{\lambda_+}\right) dr - 2 \int^{r^\dagger}_0 \sqrt{\tilde{\lambda_+}} dr} - e^{-2 \int^{r*}_0 \sqrt{\lambda_+} dr + \int^{r_M}_0 \left(\sqrt{\lambda_+}-\sqrt{\tilde{\lambda_+}}  \right) dr} \right] \right\}. \nonumber \label{eq:eplusmatchingtp} \end{eqnarray}

In either case, we now have six equations (Eqs \ref{eq:fplus}, ~\ref{eq:eminus}, ~\ref{eq:fminus} and $E_+=0$, combined with Eq. \ref{eq:matching} and/or ~\ref{eq:fplusmatchingtp} and ~\ref{eq:eplusmatchingtp}, plus the boundary conditions which fix two of the $C$ coefficients) in six unknowns ($C_1$, $C_2$, $C_3$, $C_4$, $A_+$ and $A_-$). Thus we can solve for the large-$r$ behavior of the exponential solution in terms of the boundary conditions at the origin; the results are summarized in \S \ref{sec:exactsolsfinal}.

\bibliographystyle{JHEP-2}
\bibliography{twostatesommerfeld}

\end{document}